\documentclass[times,twocolumn,final]{elsarticle}
\usepackage{cag}
\usepackage{framed,multirow}

\usepackage{amssymb}
\usepackage{latexsym}
\usepackage{makecell}
\usepackage{url}
\usepackage{xcolor}
\usepackage{graphicx}
\usepackage{amsmath,amssymb,amsfonts} 
\usepackage{algorithm}
\usepackage{algorithmicx}
\usepackage{algpseudocode}
\usepackage{textcomp}
\usepackage{amsmath}
\usepackage{array}
\usepackage{overpic}
\usepackage[T1]{fontenc}
\usepackage{float} 
\usepackage{subfigure}
\usepackage{caption}
\usepackage{wrapfig} 
\usepackage{bbding}
\usepackage{multirow}
\usepackage{bm}
\definecolor{newcolor}{rgb}{.8,.349,.1}

\usepackage{hyperref}

\usepackage[switch,pagewise]{lineno} %Required by command \linenumbers below

\journal{Computers \& Graphics}

\begin{document}

\verso{Preprint Submitted for review}

\begin{frontmatter}

\title{Multiscale Progressive Text Prompt Network for Medical Image Segmentation}%

%\received{1 February 2017}
\received{\today}
%%%% Do not use the below for submitted manuscripts
%\finalform{28 March 2017}
%\accepted{2 April 2017}
%\availableonline{15 May 2017}
%\communicated{S. Sarkar}

\author[1]{Xianjun Han}
\author[1]{Qianqian Chen}
\author[1]{Zhaoyang Xie}
\author[1]{Xuejun Li*}
\cortext[cor1]{Corresponding author: Xuejun Li;
	email: xjli@ahu.edu.cn}
%\emailauthor{example@email.com}{Corresponding Author Name}
%\ead{example@email.com}

\author[2]{Hongyu Yang}
%\fntext[fn1]{Footnote 1.}  

\address[1]{No. 111 Jiulong Road, Hefei, Anhui, China, 230601} 
\address[2]{No. 24 South Section 1, Yihuan Road, Chengdu, Sichuan, China, 610065}

\begin{abstract}
%%%
The accurate segmentation of medical images is a crucial step in obtaining reliable morphological statistics. However, training a deep neural network for this task requires a large amount of labeled data to ensure high-accuracy results. To address this issue, we propose using progressive text prompts as prior knowledge to guide the segmentation process. Our model consists of two stages. In the first stage, we perform contrastive learning on natural images to pretrain a powerful prior prompt encoder (PPE). This PPE leverages text prior prompts to generate multimodality features. In the second stage, medical image and text prior prompts are sent into the PPE inherited from the first stage to achieve the downstream medical image segmentation task. A multiscale feature fusion block (MSFF) combines the features from the PPE to produce multiscale multimodality features. These two progressive features not only bridge the semantic gap but also improve prediction accuracy. Finally, an UpAttention block refines the predicted results by merging the image and text features. This design provides a simple and accurate way to leverage multiscale progressive text prior prompts for medical image segmentation. Compared with using only images, our model achieves high-quality results with low data annotation costs. Moreover, our model not only has excellent reliability and validity on medical images but also performs well on natural images. The experimental results on different image datasets demonstrate that our model is effective and robust for image segmentation. 
%%%%
\end{abstract}


\end{frontmatter}

%\linenumbers

\section{Introduction}
The accurate delineation of lesions or anatomical structures is a crucial step in clinical intervention, treatment planning, and building a computer-aided diagnosis (CAD) system. Multiscale information fusion is a common skill needed to enhance the segmentation performance by obtaining different scale representations of medical images, which can provide more useful information. For instance, MT-UNet \cite{wang2022mixed} integrated a mixed transformer module (MTM) and convolution as basic blocks into U-Net for accurate medical image segmentation. D-former \cite{wu2022d} employed a dilated Transformer and presented a U-shaped encoder-decoder hierarchical architecture that attained high effectiveness on 3D medical image segmentation. Nonetheless, these networks generated different scale representations from the encoders and processed them sequentially in the decoders, potentially leading to insufficient multiscale information fusion. HRSTNet \cite{wei2022high} exchanged internal statistics from the different resolution feature maps and fused them as input to the next layer. By utilizing this mechanism, HRSTNet maintained high-resolution features to provide spatially precise information. Propagating both high- and low-level image features made this model obtain more efficient segmentation predictions. However, it should be noted that these networks were weak in addressing the semantic gap between the multiscale features.

To enhance the quality of medical image segmentation, multimodal fusion-based methods have been found to be effective. Examples include HyperDense-Net \cite{dolz2018hyperdense}, MMNet \cite{jiang2020multi}, and MAML \cite{zhang2021modality}. These methods provide promising results but face challenges in finding paired modal medical images due to concerns over patient data privacy and intellectual property. Additionally, processing different image data can be computationally intensive, especially for accurate segmentations that require high-quality images.

Recently, vision-language models have achieved remarkable success in natural image processing due to the fact that language information can be treated as pseudolabels. Compared to images, processing text can significantly reduce computational costs. Therefore, incorporating text as prompts has gained significant attention from both academic and industrial researchers. This has led to a question being raised about whether the knowledge learned from natural image and text pairs through large pre-trained vision-language models can benefit medical image understanding. However, there exists a domain gap that has spurred researchers to explore more powerful approaches. This may pose a challenge in generating high-quality pseudolabels. Thus, designing an excellent pretrained model and finding suitable treatments to overcome the semantic gap are still topics that warrant further study to extract preferable image features.

To achieve satisfactory visual effects, researchers have been working on designing robust encoders that can adapt to multiple visual tasks. Contrastive learning (CL) has emerged as a top method for training an encoder to produce good representations. Most CL methods are based on Siamese networks, which are tools for intrinsic comparison. They typically use the same image's two views (positive pairs) and reject different images (negative pairs). To improve this mechanism, BYOL \cite{grill2020bootstrap} trained a momentum encoder (online network) only on positive pairs to predict the target network representation. With the convergence of the model, the momentum encoder could achieve superior performance in extracting representations. Furthermore, SimSiam \cite{chen2021exploring} can be thought of as "BYOL without the momentum encoder". The stop-gradient operation prevented the collapse of the model. Inspired by this, we pretrain an encoder on natural image-text pairs using text prior prompts to extract meaningful representations. This approach is rarely used to pretrain medical image segmentation networks.

To ensure the quality of pseudolabels, Transformer was used in natural language processing (NLP) to build global connections between sequence tokens. This approach has been extended to computer vision and proposed as Vision Transformer (ViT \cite{dosovitskiy2020image}) to overcome the lack of inductive bias restriction. ViLT \cite{kim2021vilt} used a simple linear projection without convolutional neural networks (CNN) to obtain text statistics. However, Kim \emph{et al.} \cite{kim2021vilt} assumed that multimodal interaction was cardinal instead of emphasizing the embedding of image and text. MMMAE \cite{chen2022multi} was based on a pure Transformer to reconstruct medical image tokens and terminology. These methods generally overlooked the local feature extraction capability of CNNs. Additionally, the feature maps generated by ViT were single-scale and low-resolution, leading to suboptimal performances in medical image segmentation tasks. Recently, SAM \cite{kirillov2023segment} has gained widespread attention for its ability to apply zero-shot learning to new image segmentation tasks. However, Ji et al. \cite{ji2023segment} have discovered that SAM requires substantial human prior knowledge for medical image tasks. Furthermore, experimental results indicate that SAM may encounter difficulties in accurately identifying amorphous lesion regions in various anatomical structures.

To address the challenges mentioned above, our language-driven model incorporates text prior prompts to guide downstream segmentation tasks. This enables our model to capture context semantics and generate high-quality masks at a lightweight computational cost. We propose a two-stage training pipeline to extract meaningful progressive features. In the first stage, PPE is trained on natural data using contrastive learning. In the second stage, the PPE inherited from the first stage generates single-scale multimodality features. MSFF receives these features and transforms them into multiscale multimodality features while merging text and image information. This progressive method not only effectively addresses the semantic gap between natural data and medical data but also improves the accuracy of prediction masks. UpAttention is used to refine the features from MSFF, which is essential for accurate masks. To obtain a useful representation and achieve a superior visual effect, we combine the U-shaped Transformer and CNN to form the basic block of PPE. It is trained using contrastive learning in the first stage, which enables it to extract meaningful representations effectively. Additionally, due to the abundance of natural images with well-labeled text depictions, we can alleviate the stress caused by scarce medical data.

\begin{figure*}[tbp]
	\centering
	%\hfill
	%\includegraphics[width=0.98\textwidth]{fig/model1.png}
	\includegraphics[height=6cm,width=12.5cm]{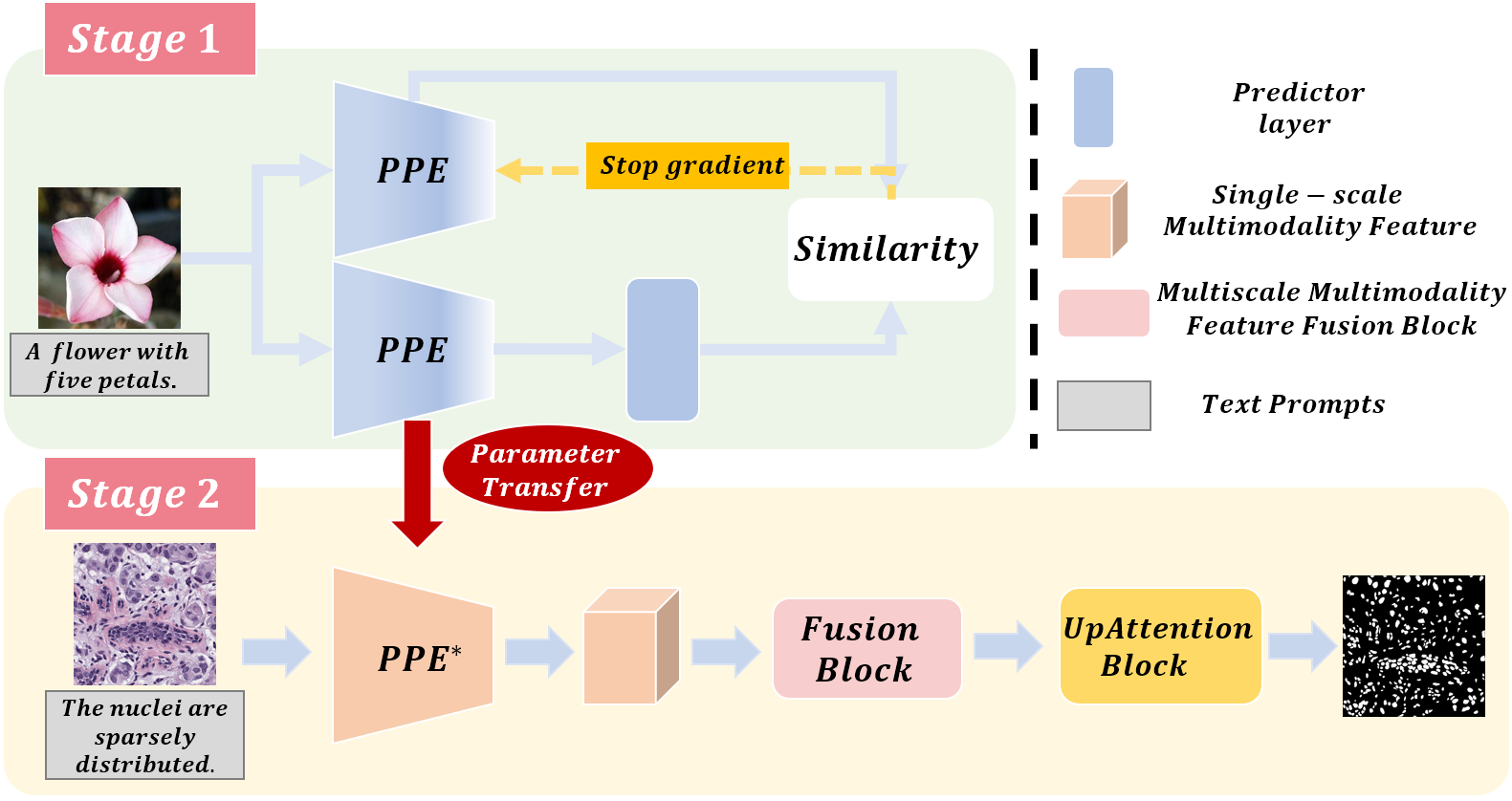}
	\caption{\label{fig.model}%
		\small The overview of network framework. \textbf{Stage 1}: $PPE$ is trained using contrastive learning to acquire the ability to capture image and text representations. \textbf{Stage 2}: $PPE^{*}$ inherits the network parameters of $PPE$.
		We use $PPE^{*}$ to generate the multimodality features and send them into the MSFF, which can extract multiscale multimodality features. The UpAttention block refines the predicted masks.}
	\vspace*{-0.3cm}
\end{figure*}

In brief, the contribution of the proposed method can be summarized as follows:
\begin{enumerate}
	\item We present multiscale features of medical images for segmentation, which can capture more valuable information. Furthermore, we make them progressive by transforming single-scale features into multiscale representations.
	\item We have introduced multimodal data to enhance medical image segmentation, which incorporates both text prior prompts and image information. Moreover, we have devised a two-stage learning pipeline that comprises pretraining and downstream tasks.
	\item We conducted a series of experiments on both medical and natural images, and the results revealed that our model is capable of capturing context semantics and producing high-quality masks.
\end{enumerate}

\section{Related Work}
\subsection{Multimodal Medical Image Segmentation}
Multimodality fusion has a profound influence on medical image segmentation because this method remedies the defect  of having only one type of statistic about a target. A host of strategies for multimodality fusion have been developed with success. The majority of modalities in medical image segmentation are computed tomography (CT), magnetic resonance imaging (MRI) and positron emission tomography (PET) \cite{zhou2019review}. For example, Guo \emph{et al.} \cite{guo2019deep} implemented medical image segmentation based on CNN with these three modalities. They designed three fusion strategies and proved that fusion within the network (at the convolutional layer or fully connected layer) was better than the fusion outside the network (at network output through voting). Therefore, there were layer-level fusion networks that could effectively integrate multimodal images. HyperDense-Net \cite{dolz2018hyperdense} uses dense connections to facilitate learning, and it brought huge improvements to many segmentation networks. Nevertheless, layer-level fusion networks cannot fully leverage and achieve advantages between modalities. Therefore, researchers have exploited fusion segmentation networks based on the decision level. Decision-level fusion networks could independently learn the complementary information from different modalities, which enhanced the segmentation performance. Most of these networks operate with averaging and majority voting. EMMA \cite{kamnitsas2017ensembles} combined a heterogeneous collection of CNNs and then averaged the confidence of the individual architectures. The voxel with the highest confidence was deemed the final segmentation.

However, these networks have trouble finding different modalities of medical images in pairs because of privacy and intellectual property. Meanwhile, processing different image data will lead to excessive computation, especially for accurate segmentation that demands  higher quality images. Therefore, researchers have emphasized language-driven medical image segmentation. 

CLIP-Driven Universal Model \cite{liu2023clip}, this work primarily focused on preprocessing text and images by incorporating embedding learned from contrastive language-image pre-training (CLIP) into segmentation models, which achieved the highest caliber results on BTCV. LViT \cite{li2022lvit}  utilized medical text annotations to compensate for the inadequacy of images. The model has paid significant attention to fusing image and language features, providing novel insights into medical image segmentation. In this paper, we aim to achieve more accurate results and build a more flexible network structure by pretraining a powerful prior prompt encoder and applying it in the downstream segmentation task.

\begin{figure*}[!h]
	\centering
	\includegraphics[width=0.8\linewidth]{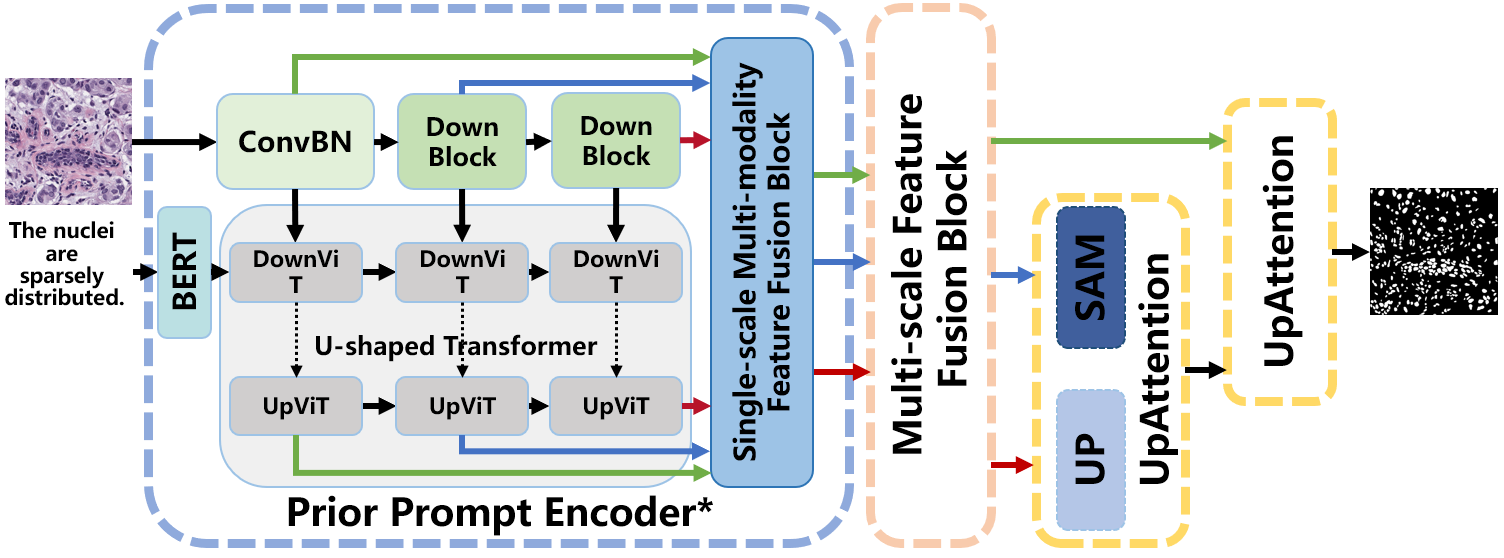}
	\caption{\small The pipeline of \textbf{Stage 2}.}
	\label{fig.stage2}
	\vspace{-0.3cm}
\end{figure*}

\subsection{Contrastive Learning }
Contrastive learning (CL) groups similar samples (positive) closer and distinguishes dissimilar pairs (negative) \cite{kocaman2022saliency}. SimCLR \cite{chen2020simple} was simple and required neither specialized networks nor a memory bank. It learned representations by maximizing the agreement between two augmented views of one data point.
MoCo \cite{he2020momentum} built a dynamic dictionary with a queue and a moving-averaged encoder. The encoder was trained by matching an encoded query to a dictionary of encoded keys, which made MoCo learn better visual representations. SwAV \cite{caron2020unsupervised} incorporated online clustering into Siamese networks without computing pairwise comparisons. 

However, such strategies only learned global representations, which are not suitable for the pixel-level segmentation task. 
Chaitanya \emph{et al.} \cite{chaitanya2020contrastive} proposed new pretrained domain-specific contrasting strategies for medical images. They used them to learn representations for the downstream segmentation task.
Inspired by this, IDEAL \cite{basak2022ideal} took a cue from it and proposed a novel CL-based medical image segmentation framework, which made some enhancements. This paper uses a simple convolutional projection head to obtain dense pixel-level features.

We are prompted by these works, and attempt to improve the performance. SimSiam \cite{chen2021exploring}, an ameliorated CL tactic, is employed as a basic training strategy to pretrain the PPE. In addition, to balance the global and local features, we integrate Transformer and CNN into PPE. In addition, we use the copious natural image dataset to help further in extracting meaningful representations. 

\subsection{Pretrained Networks in Medical Image Segmentation}
As is known to all, the more training data there is, the better the network performance we will obtain. Chen \emph{et al.} \cite{chen2019med3d} proved that pretrained models based on large-scale datasets had a great ability to extract useful general features. Moreover, a few studies \cite{erhan2009difficulty,yanai2015food,he2019rethinking} all demonstrated that pretrained models helped accelerate the training convergence and present satisfactory visual effects. 
Med3d \cite{chen2019med3d} established a general 3D backbone network that could be used for other medical downstream tasks to gain better performance than training from scratch. A major hurdle was the tedious process of gaining medical data. Swin UNETR \cite{tang2022self} trained a Transformer-based encoder as a powerful feature extractor for 3D medical image segmentation, which was a downstream application and a 1D sequence-to-sequence prediction task. Based on this, UNetFormer \cite{hatamizadeh2022unetformer} constructed a unified architecture using a 3D Swin Transformer as the encoder and hybrid transformer-based decoders. Furthermore, a self-supervised scheme was employed in the pretraining process, and its effectiveness was validated.

Despite the prevalence of image-only inputs in current methods for medical image segmentation, achieving high-accuracy results is crucial, particularly when it comes to microanatomical structure. To address this challenge, we propose using text prior prompts as pseudolabels to guide high-quality segmentation. These text annotations enable our model to better understand contextual semantics, ultimately leading to improved performance.

Moreover, relying solely on Transformers may result in an inefficient use of computing resources. Therefore, we have developed a new model that balances accuracy and speed for medical image segmentation. By incorporating these changes, we can ensure that our approach not only achieves superior results but also operates efficiently within resource limitations.

\subsection{CNN and Transformer in Medical Image Segmentation}
CNNs have achieved tremendous success in medical image segmentation. Since FCN \cite{long2015fully} was proposed, CNN has drawn extensive attention in the segmentation area. Then, U-Net \cite{ronneberger2015u} laid the foundation for using CNN for medical image segmentation. Numerous network architectures are inspired by U-Net, which is a breakthrough in this area. Although it generates great effects, CNN still has trouble in capturing the long-range dependencies between entities. 

Therefore, to remedy this CNN defect, Transformer was introduced in \cite{vaswani2017attention} and then applied to medical image analysis. Themyr \emph{et al.} \cite{themyr2022memory} designed the FINE Transformer, which could learn memory tokens to indirectly model full range interactions while controlling the memory and computational costs. TMSS \cite{saeed2022tmss} leveraged the superiority of Transformer to handle different modalities. Despite being feasibly designed, utilizing a pure Transformer for medical image segmentation might result in a limited localization capacity because of inadequate low-level features.

To better merge information from different distances, a variety of hybrid CNN-Transformer structures were used to encode global features with Transformer while retaining the CNN's ability to extract local features. CS-Unet \cite{liu2022optimizing} contains the well-designed Convolutional Swin Transformer (CST) block. It integrated convolutions and attention to obtain inherent localized spatial context and inductive biases. CATS \cite{li2022cats} is a U-shaped CNN augmented with an independent Transformer encoder model. CATS also fused the information from the convolutional encoder and the Transformer. Considering the high computational cost, the hybrid CNN-Transformer structure and the type of input are still topics worthy of further study.

\section{Method}
\subsection{Overview of Network Architecture}
Our model incorporates a prior prompt encoder (PPE), followed by a multiscale feature fusion block and an UpAttention block, to achieve accurate medical image segmentation. We have innovatively introduced text prior prompts to guide the delineation of some tiny segmentation areas. This contributes to fusing semantic features from various data formats into a single structure, while also reducing computation compared to using different image modalities.

\begin{algorithm}[!h]
	\caption{Multiscale Progressive Text Prompt Network} %算法的名字
	\label{alg}
	\hspace*{0.02in} {\bf Input:} %算法的输入， \hspace*{0.02in}用来控制位置，同时利用 \\ 进行换行
	$\enspace X_{img}$: original images;\\
	\hspace*{0.5in}$\enspace X_{text}$: corresponding text annotations;\\
	\hspace*{0.02in} {\bf Output:} %算法的结果输出
	$Y$: the segmented masks;
	\begin{algorithmic}[1]
		\State \emph{Stage 1 : training based on contrastive learning}
		\For{each training iteration} % For 语句，需要和EndFor对应
		\State $X_{1}, X_{2} \leftarrow aug(X_{img}), aug(X_{img})$\textcolor[rgb]{0.25, 0.5, 0.75}{\# random augmentation}
		\State $Y_{1}, Y_{2} \leftarrow \mathcal{F}(X_{1}, X_{text}), \mathcal{F}(X_{2}, X_{text})$\textcolor[rgb]{0.25, 0.5, 0.75}{\# $\mathcal{F}$ denotes the PPE}
		\State $P_{1}, P_{2} \leftarrow \mathcal{G}(Y_{1}), \mathcal{G}(Y_{2})$\textcolor[rgb]{0.25, 0.5, 0.75}{\# predictions}
		\State 
  $\mathcal{L} \leftarrow \mathcal{D}(P_{1}, Y_{2}) + \mathcal{D}(P_{2}, Y_{1})
  $\textcolor[rgb]{0.25, 0.5, 0.75}{\# $\mathcal{D}$ is the negative cosine similarity, and the second parameter stops gradient.  }
		\State $\mathcal{L}$.backward() \textcolor[rgb]{0.25, 0.5, 0.75}{\# back-propagate}
		\State update($\mathcal{F}$, $\mathcal{G}$) \textcolor[rgb]{0.25, 0.5, 0.75}{\# SGD update}
		\EndFor
		
		\State \emph{Stage 2 : Multiscale Multimodality Feature Fusion and Refining}
		\For{each training iteration}
		\State \textcolor[rgb]{0.25, 0.5, 0.75}{\# $PPE^{*}$ inherits the parameters of $\mathcal{F}$ from $Stage 1$.}
		\State $y_1, y_2, y_3 \leftarrow PPE^{*}(X_{img}, X_{text})$;
		\State \textcolor[rgb]{0.25, 0.5, 0.75}{\# $\hat{y}_1, \hat{y}_2, \hat{y}_3$ are multiscale representations.}
		\State $\hat{y}_1, \hat{y}_2, \hat{y}_3 \leftarrow MSFF(y_1, y_2, y_3)$;
		\State $Y \leftarrow  UpAttention(\hat{y}_1, UpAttention(\hat{y}_2, \hat{y}_3)) $;
		\EndFor
		\State \Return $Y$
	\end{algorithmic}
	\vspace{-0.09cm}
\end{algorithm}

As shown in Fig. \ref{fig.model}, our model consists of two stages. In the first stage, we utilize natural text prior prompts to pretrain $PPE$ using contrastive learning. Two identical $PPE$ models are trained in separate pipelines, with the goal of maximizing feature similarity between them. By applying stop gradient operation on one $PPE$ model, we ensure convergence of the other $PPE$ model. Additionally, we leverage a large number of natural images with well-annotated text prior prompts to alleviate the burden of sparse medical datasets. $PPE$ reconciles these two modalities to produce single-scale multimodality features, which are further enhanced by integrating CNN and Transformer modules into $PPE$ to balance local and global features.

Fig. \ref{fig.stage2} shows the second stage and $PPE^{*}$ inherits the network parameters of $PPE$. To exchange contextual information from the single-scale multimodality features generated by $PPE^{*}$, we have designed a multiscale feature fusion block (MSFF) and an UpAttention block. The MSFF module combines patch merging and expanding processes to obtain richer semantic information, generating multiscale and multimodal features by merging different feature maps of varying sizes and integrating textual information. These two progressive features help bridge the semantic gap between natural data and medical data. The two smallest scales of the outputs from MSFF will be processed by the UpAttention block, which includes global average pooling, global max pooling, and upsampling operations. Another UpAttention block receives the output of the first UpAttention block and the largest feature from MSFF. Finally, we obtain the segmented result. The corresponding pseudo code is presented in Algorithm \ref{alg}.

\subsection{Stage 1: Single-scale Multimodality Features via Contrastive Learning}
\subsubsection{Contrastive Learning} As shown in Fig. \ref{fig.model}, $Stage 1$ uses contrastive learning as the training strategy. First, we use text as prior prompts. The input image is augmented randomly into two new views. It is worth mentioning that the augmentation operations do not include flip or crop because some text annotations contain position information. Second, two views will be processed by the same encoder $PPE$ networks. The predictor layer and the stop gradient operation are applied on different sides. $PPE$ with the predictor layer will gradually converge by maximizing the feature similarity between the two branches. Finally, $PPE^{*}$ inherits the network parameters of the convergent $PPE$ to generate single-scale multimodality features. Because this strategy demands massive amounts of data to ensure model convergence, copious amounts of natural data are applied to enhance the performance of $PPE$. This initial stage's processing not only alleviates the burden of training data but also accelerates the convergence rate of the subsequent stage.

\begin{figure}[!h]
	\centering
	\includegraphics[width=0.7\linewidth]{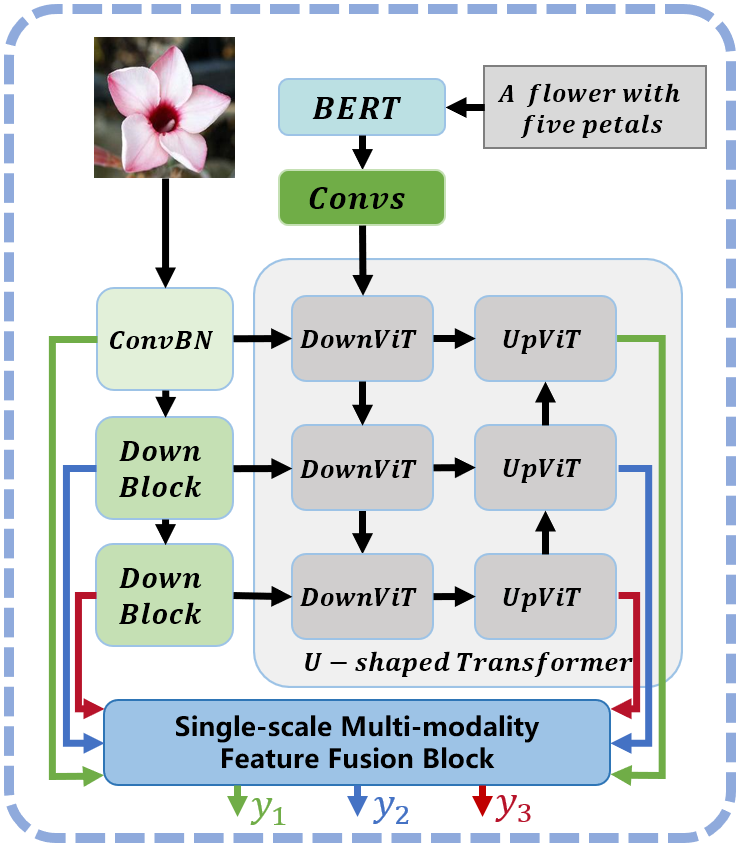}
	\caption{\small The network architecture of PPE. (BERT is a trained language model that can be used to extract text features. The U-shaped Transformer includes DownViT and UpViT, which are used to fuse features.)}
	\label{fig.PTE}
	\vspace{-0.3cm}
\end{figure}

\subsubsection{PPE} PPE is divided into the image-input branch and the text-input branch. 
The image-input branch contains a convolution block and two downsampling blocks that extract image features from different levels. Then, the outputs are sent into the U-shaped Transformer in the text-input branch.
The text-input branch is comprised of BERT \cite{devlin2018bert} and the U-shaped Transformer, where we construct 3 layers of ViT based on U-Net. BERT helps obtain text representations with a bidirectional encoder from the text prior prompts. The U-shaped Transformer receives image features and text representations for processing. Such a design helps PPE fuse more meaningful representations. 

As shown in Fig. \ref{fig.PTE}, in the image-input piece, the input images $X_{img}$ are first processed by the ConvBN block, including the convolution, BatchNorm (BN) and ReLU activation layers. This block helps extract shallow features $X_1$. Second, the DownBlock comprises the ConvBN block and the max pooling layer. We stack two DownBlocks to downsample the feature successively.
\begin{equation}
	\begin{aligned}
		&ConvBN = ReLU(BN(Conv2d()))\\
		&DownBlock = MaxPooling(ConvBN())\\
		&X_1 = ConvBN(X_{img}) \\ 
		&X_{i+1} = DownBlock(X_i), \quad i = 1,2
	\end{aligned}	
	\label{imageBranch}	
\end{equation}

In the text-input piece, BERT is used to convert a single word into a 768-dimensional word vector $X_{text}$. Then, 4 convolution layers are employed to process the preliminary process text vectors. 
\begin{equation}
	\begin{aligned}
		&X_{text,4} = Conv1d(X_{text})\\
		&X_{text,i-1} = Conv1d(X_{text,i}), \quad i = 4,3,2
	\end{aligned}	
	\label{text}	
\end{equation}

\begin{figure}[!h]
	\centering
	\includegraphics[width=0.9\linewidth]{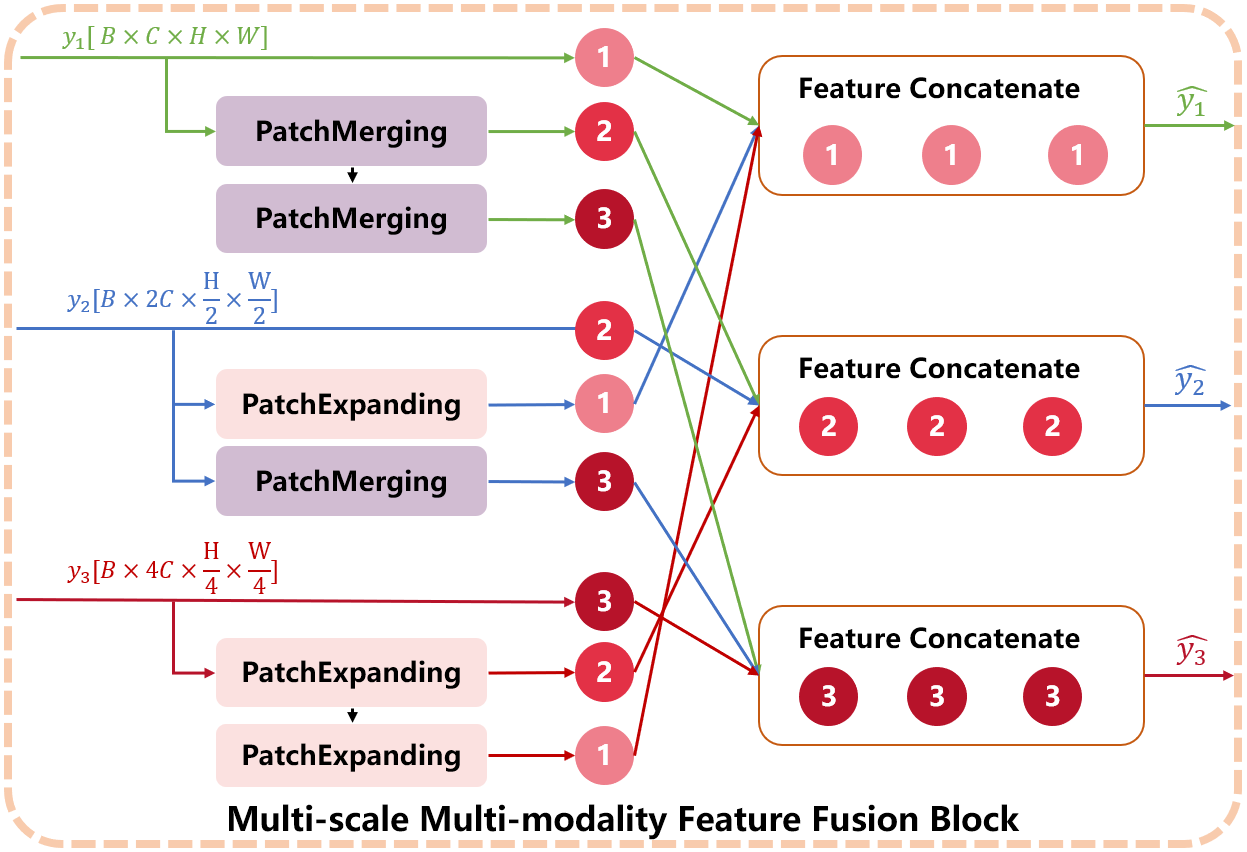}
	\caption{\small The network architecture of MSFF.}
	\label{fig.MSFF}
	\vspace{-0.3cm}
\end{figure}

\begin{figure*}[!h]
	\centering
	\includegraphics[width=0.8\textwidth]{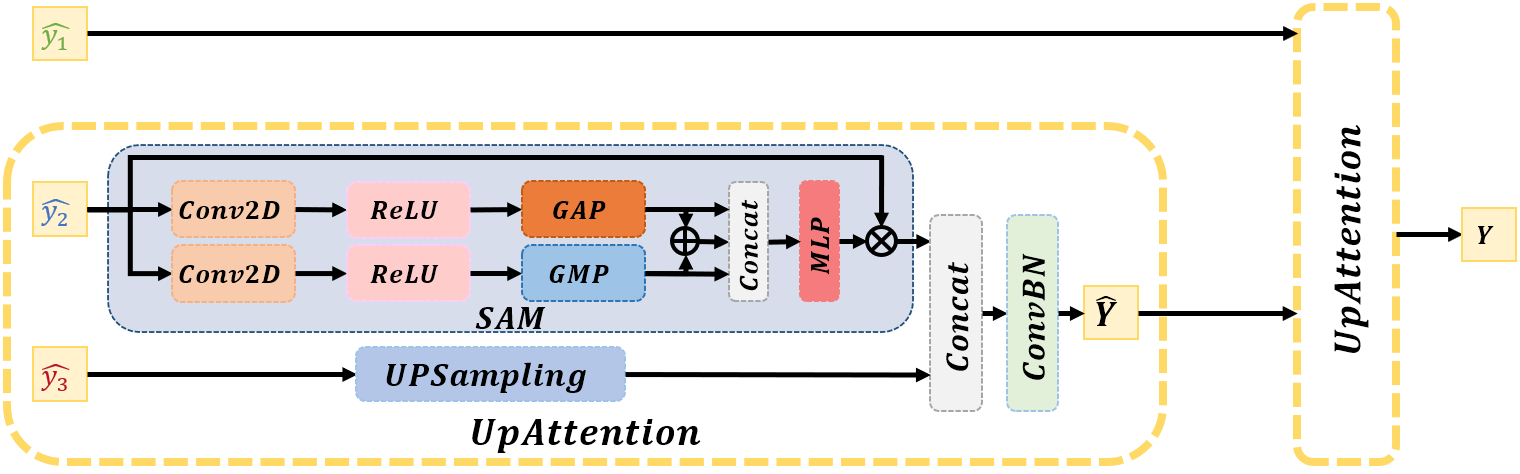}
	\caption{\small The network architecture of UpAttention. (SAM denotes the spatial attention module. GAP and GMP represent the global average pooling and global max pooling, respectively.)}
	\label{fig.UpAttention}
	\vspace{-0.3cm}
\end{figure*}

The U-shaped Transformer contains 3 layers, and downup ViTs form each layer. In the first DownViT, a PatchEmbedding layer is used to reshape the image tensors $X_1$ into sequences of flattened 2D patches $X_{p,1}$. Then, $X_{text,1}$ will be processed by a ConvBN block and added to $X_{p,1}$. This method helps further reconcile the image and text information. Finally, the output will go through the Transformer part.
\begin{equation}
	\begin{aligned}
		&ConvBN = ReLU(BN(Conv1d()))\\
		&X_{p,1} = PatchEmbedding(X_{1})\\
		&Y = X_{p,1} + ConvBN(X_{text,1})\\
		&Y^{'} = MSA(LN(Y)) + Y\\
		&Y_{1}= MLP(LN(Y^{'})) + Y^{'}
	\end{aligned}	
	\label{DownViT1}	
\end{equation}
where $X_{text,1}$ is from \eqref{text}.

The second and third DownViT do not need to process the text features. They are responsible for fusing the output $Y_i$ from Transformer and the image features $X_i$.
\begin{equation}
	\begin{aligned}
		&ConvBN = ReLU(BN(Conv1d()))\\
		&X_{p,i+1} = PatchEmbedding(X_{i+1})\\
		&Y^{'}_{i+1} = MSA(LN(X_{p,i+1})) + X_{p,i+1}\\
		&Y_{i+1} = MLP(LN(Y^{'}_{i+1})) + Y^{'}_{i+1}\\
		&Y_{i+1} = ConvBN(Y_{i+1})\\
		&Y_{i+1} = Concat(Y_{i+1}, Y_{i}), \quad i = 1, 2\\
	\end{aligned}	
	\label{DownViT23}	
\end{equation}
where $X_{i+1}$ is from \eqref{imageBranch} and $Concat$ represents the concatenate function in PyTorch. $MSA$ denotes the multihead self-attention block, which is an extension of self-attention (SA). It runs several SA operations in parallel, called "heads", and projects their concatenated outputs \cite{dosovitskiy2020image}. $MLP$ represents the multilayer perceptron, including the linear function, GeLU activation and dropout function.

UpViT is similar to DownViT. The third bottom UpViT only consists of the Transformer part.
\begin{equation}
	\begin{aligned}
		&Y^{'}_{3} = MSA(LN(Y_{3})) + Y_{3}\\
		&Y_{3} = MLP(LN(Y^{'}_{3})) + Y^{'}_{3}\\
	\end{aligned}	
	\label{UpViT3}	
\end{equation}
The input $Y_{3}$ is calculated from \eqref{DownViT23}.

The second and first UpViT are in charge of reconciling the output from the corresponding DownViT and new $Y_{i}$.
\begin{equation}
	\begin{aligned}
		&ConvBN = ReLU(BN(Conv1d()))\\
		&Y^{'}_{i} = MSA(LN(Y_{i})) + Y_{i}\\
		&Y_{i} = MLP(LN(Y^{'}_{i})) + Y^{'}_{i}\\
		&Y_{i+1} = ConvBN(Y_{i+1})\\
		&Y_{i} = Y_{i} + Y_{i+1}, \quad i = 1, 2
	\end{aligned}	
	\label{UpViT12}	
\end{equation}

The single-scale multimodality feature fusion block is applied to further fuse the image and text statistics and correct the dim of the results. First, we will reshape $Y_i \in B \times N_{patch} \times C$ into $Y_i \in B \times C \times H \times W$. Here, $B$ represents a batch, $C$ represents the channel of the input tensor, and $N_{patch}$ represents the number of patches. $H$ and $W$ denote the height and width of an image, respectively.
Then, $Y_i$ will be upsampled. The convolution, BN and ReLU activation layers are the subsequent operations. Finally, $Y_i$ is added to $X_i$.
\begin{equation}
	\begin{aligned}
		&ConvBN = ReLU(BN(Conv2d()))\\
		&Y_{i} =  ConvBN(UpSample(reshape(Y_{i})))\\
		%&Y_{i} = Upsample(Y_{i})\\
		%&Y_{i} = ConvBN(Y_{i})\\
		&Y_{i} = Y_{i} + X_{i}, \quad i = 1, 2, 3
	\end{aligned}	
	\label{reconstruction}	
\end{equation}
After this block, we obtain 3 groups of single-scale multimodality features.

\subsection{Stage 2: Multiscale Multimodality Feature Fusion and Refining}
\subsubsection{Multiscale Feature Fusion Block}
As shown in Fig. \ref{fig.MSFF}, this block is employed to generate the progressive features, which helps compensate for the semantic gap between the natural data and medical data. The single-scale medical image features from PPE are $B \times C \times H \times W$, $B \times 2C \times  \frac{H}{2} \times \frac{W}{2}$ and $B \times 4C \times  \frac{H}{4} \times \frac{W}{4}$, denoted as $tensor_1$, $tensor_2$, and $tensor_3$, respectively. They will be developed into multiscale features via this block. The channels, height and width of images are various. This enables adjacent scales of context features to be more precisely incorporated.

We give a detailed description of the patch merging block and the patch expanding block.
Taking $B \times C \times H \times W$ as an example, the patch merging block reshapes it into $B \times H \times W \times C$. Then, we join adjacent $2 \times 2$ patches via a slice operation, which yields four tensors with $B \times \frac{H}{2} \times \frac{W}{2} \times C$. These four are concatenated according to the dim of $C$, so we obtain a tensor with $B \times \frac{H}{2} \times \frac{W}{2} \times 4C$. After that, the output is processed by the layernorm function, and the channel $4C$ is lowered to $2C$ by the linear layer. Finally, we permute the result into $B \times 2C \times  \frac{H}{2} \times \frac{W}{2}$.

The patch expanding block also reshapes $B \times C \times H \times W$ into $B \times H \times W \times C$. Channel $C$ can disassemble into $2 \times 2 \times c$, where $c = \frac{C}{4}$. Then, the rearrange function will turn the input into $B \times 2H \times 2W \times \frac{C}{4}$. Finally, the layernorm, linear and permute operations are the same as the patch merging block.

We stack two patch merging blocks to process multiscale features, so we obtain a feature group $\left\{B \times C \times H \times W, B \times 2C \times \frac{H}{2} \times \frac{W}{2}, B \times 4C \times \frac{H}{4} \times \frac{W}{4}\right\}$. For $tensor_{2}$, a patch expanding block and a patch merging block are applied to amplify and narrow it. Two patch expanding blocks are stacked to magnify $tensor_{3}$. In addition, these three feature groups have the same size features as the group above, which are also $\left\{B \times C \times H \times W, B \times 2C \times \frac{H}{2} \times \frac{W}{2}, B \times 4C \times \frac{H}{4} \times \frac{W}{4}\right\}$.
Since obtaining 3 feature groups, we select the same size features in these groups. Specifically, three $B \times C \times H \times W$ from 3 groups are concatenated according to the dim of the channel $C$, so we obtain $B \times 3C \times H \times W$. The identical operation is performed on other sizes of tensors. As a result, new $tensor_1$, $tensor_2$, and $tensor_3$ will be $B \times 3C \times H \times W$, $B \times 6C \times  \frac{H}{2} \times \frac{W}{2}$ and $B \times 12C \times  \frac{H}{4} \times \frac{W}{4}$.
Multiscale multimodality features are an extension of the features from PPE. This progressive operation remedies the defect of the semantic gap between the natural data and medical data.

\begin{figure*}[h!]
	\centering
	\begin{minipage}{0.05\linewidth}
		\centering
		\centerline{{\scriptsize \makecell{The nuclei are\\ sparsely\\ distributed.}}}
	\end{minipage}
	\begin{minipage}{0.11\linewidth}
		\centering
		\includegraphics[width=0.55in,height=0.55in]{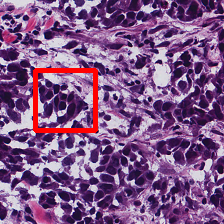}
		\label{IMG1}
	\end{minipage}
	\begin{minipage}{0.11\linewidth}
		\centering
		\includegraphics[width=0.55in,height=0.55in]{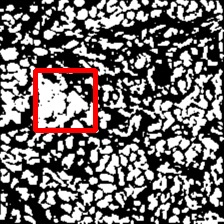}
		\label{MedT1}
	\end{minipage}
	\begin{minipage}{0.11\linewidth}
		\centering
		\includegraphics[width=0.55in,height=0.55in]{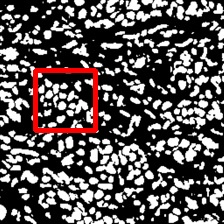}
		\label{GTUNet1}
	\end{minipage}
	\begin{minipage}{0.11\linewidth}
		\centering
		\includegraphics[width=0.55in,height=0.55in]{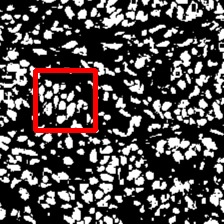}
		\label{SwinUNet1}
	\end{minipage}
	\begin{minipage}{0.11\linewidth}
		\centering
		\includegraphics[width=0.55in,height=0.55in]{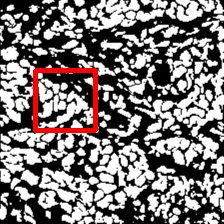}
		\label{UCTransNet1}
	\end{minipage}
	\begin{minipage}{0.11\linewidth}
		\centering
		\includegraphics[width=0.55in,height=0.55in]{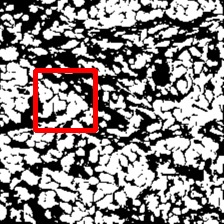}
		\label{LViT1}
	\end{minipage}
	\begin{minipage}{0.11\linewidth}
		\centering
		\includegraphics[width=0.55in,height=0.55in]{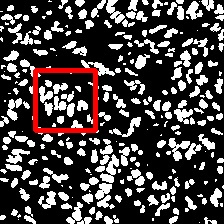}
		\label{OURS}
	\end{minipage}
	\begin{minipage}{0.11\linewidth}
		\centering
		\includegraphics[width=0.55in,height=0.55in]{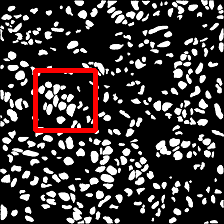}
		\label{GT1}
	\end{minipage}
	
	\vspace{0.1cm}
	\begin{minipage}{0.05\linewidth}
		\centering
		\centerline{{\scriptsize \makecell{The nuclei are\\ sparsely\\ distributed.}}}
	\end{minipage}
	\begin{minipage}{0.11\linewidth}
		\centering
		\includegraphics[width=0.55in,height=0.55in]{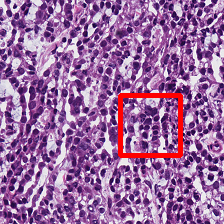}
		\label{IMG2}
	\end{minipage}
	\begin{minipage}{0.11\linewidth}
		\centering
		\includegraphics[width=0.55in,height=0.55in]{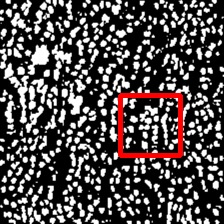}
		\label{MedT2}
	\end{minipage}
	\begin{minipage}{0.11\linewidth}
		\centering
		\includegraphics[width=0.55in,height=0.55in]{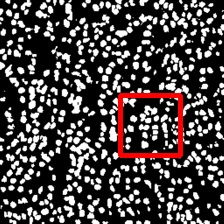}
		\label{GTUNet2}
	\end{minipage}
	\begin{minipage}{0.11\linewidth}
		\centering
		\includegraphics[width=0.55in,height=0.55in]{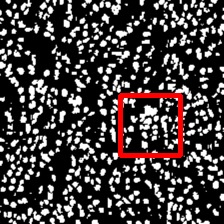}
		\label{SwinUNet2}
	\end{minipage}
	\begin{minipage}{0.11\linewidth}
		\centering
		\includegraphics[width=0.55in,height=0.55in]{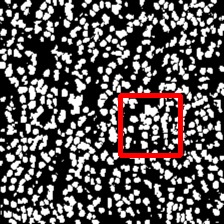}
		\label{UCTransNet2}
	\end{minipage}
	\begin{minipage}{0.11\linewidth}
		\centering
		\includegraphics[width=0.55in,height=0.55in]{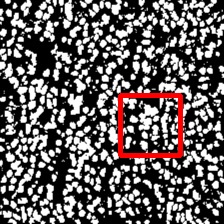}
		\label{LViT2}
	\end{minipage}
	\begin{minipage}{0.11\linewidth}
		\centering
		\includegraphics[width=0.55in,height=0.55in]{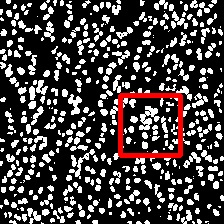}
		\label{OURS2}
	\end{minipage}
	\begin{minipage}{0.11\linewidth}
		\centering
		\includegraphics[width=0.55in,height=0.55in]{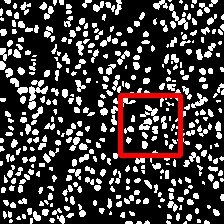}
		\label{GT2}
	\end{minipage}
	
	\vspace{0.1cm}
	\hspace{-0.2cm}
	\begin{minipage}{0.05\linewidth}
		\vspace{-0.7cm}
		\centering
		\centerline{{\scriptsize \makecell{The nuclei\\ density in the\\ lower left\\ is high.}}}
	\end{minipage}
	\begin{minipage}{0.11\linewidth}
		\centering
		\includegraphics[width=0.55in,height=0.55in]{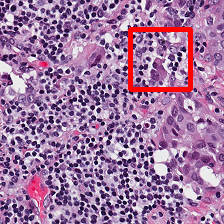}
		\centerline{\scriptsize  IMG}
		\label{IMG3}
	\end{minipage}
	\begin{minipage}{0.11\linewidth}
		\centering
		\includegraphics[width=0.55in,height=0.55in]{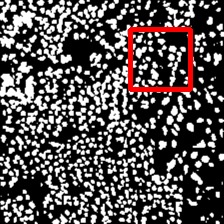}
		\centerline{\scriptsize  MedT}
		\label{MedT3}
	\end{minipage}
	\begin{minipage}{0.11\linewidth}
		\centering
		\includegraphics[width=0.55in,height=0.55in]{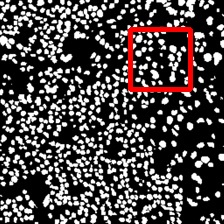}
		\centerline{\scriptsize  GTUNet}
		\label{GTUNet3}
	\end{minipage}
	\begin{minipage}{0.11\linewidth}
		\centering
		\includegraphics[width=0.55in,height=0.55in]{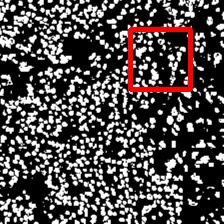}
		\centerline{\scriptsize  SwinUNet}
		\label{SwinUNet3}
	\end{minipage}
	\begin{minipage}{0.11\linewidth}
		\centering
		\includegraphics[width=0.55in,height=0.55in]{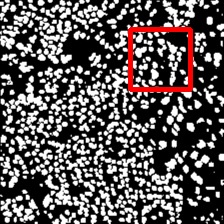}
		\centerline{\scriptsize  UCTransNet}
		\label{UCTransNet3}
	\end{minipage}
	\begin{minipage}{0.11\linewidth}
		\centering
		\includegraphics[width=0.55in,height=0.55in]{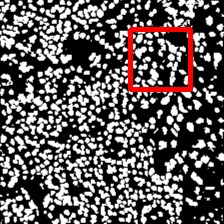}
		\centerline{\scriptsize  LViT}
		\label{LViT3}
	\end{minipage}
	\begin{minipage}{0.11\linewidth}
		\centering
		\includegraphics[width=0.55in,height=0.55in]{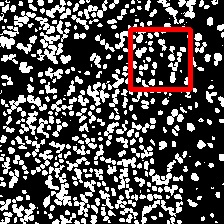}
		\centerline{\scriptsize  OURS}
		\label{OURS3}
	\end{minipage}
	\begin{minipage}{0.11\linewidth}
		\centering
		\includegraphics[width=0.55in,height=0.55in]{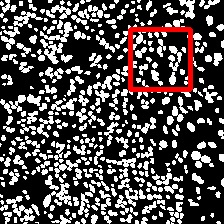}
		\centerline{\scriptsize  GT}
		\label{GT3}
	\end{minipage}
	\caption{Medical image segmentation from MoNuSeg dataset. The text annotations are provided by LViT \cite{li2022lvit}.}
	\label{test1}
\end{figure*}
\begin{figure*}[h!]
	\centering
	\begin{minipage}{0.05\linewidth}
		\centering
		\centerline{{\scriptsize \makecell{The two\\ infected regions\\ are symmetric.}}}
	\end{minipage}
	\begin{minipage}{0.11\linewidth}
		\centering
		\includegraphics[width=0.55in,height=0.55in]{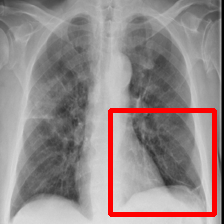}
		\label{IMG4}
	\end{minipage}
	\begin{minipage}{0.11\linewidth}
		\centering
		\includegraphics[width=0.55in,height=0.55in]{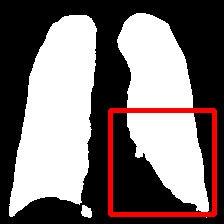}
		\label{MedT4}
	\end{minipage}
	\begin{minipage}{0.11\linewidth}
		\centering
		\includegraphics[width=0.55in,height=0.55in]{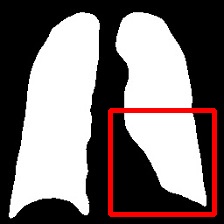}
		\label{GTUNet4}
	\end{minipage}
	\begin{minipage}{0.11\linewidth}
		\centering
		\includegraphics[width=0.55in,height=0.55in]{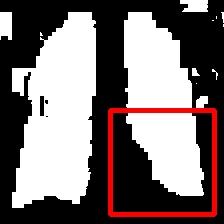}
		\label{SwinUNet4}
	\end{minipage}
	\begin{minipage}{0.11\linewidth}
		\centering
		\includegraphics[width=0.55in,height=0.55in]{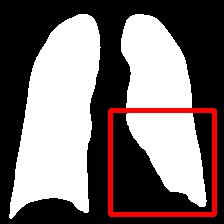}
		\label{UCTransNet4}
	\end{minipage}
	\begin{minipage}{0.11\linewidth}
		\centering
		\includegraphics[width=0.55in,height=0.55in]{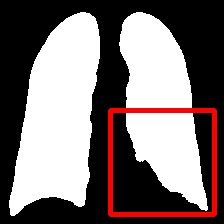}
		\label{LViT4}
	\end{minipage}
	\begin{minipage}{0.11\linewidth}
		\centering
		\includegraphics[width=0.55in,height=0.55in]{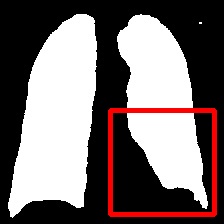}
		\label{OURS4}
	\end{minipage}
	\begin{minipage}{0.11\linewidth}
		\centering
		\includegraphics[width=0.55in,height=0.55in]{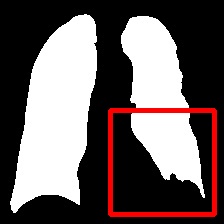}
		\label{GT4}
	\end{minipage}
	
	\vspace{0.1cm}
	\begin{minipage}{0.05\linewidth}
		\centering
		\centerline{{\scriptsize \makecell{The two\\ infected regions\\ are symmetric.}}}
	\end{minipage}
	\begin{minipage}{0.11\linewidth}
		\centering
		\includegraphics[width=0.55in,height=0.55in]{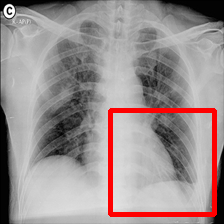}
		\label{IMG5}
	\end{minipage}
	\begin{minipage}{0.11\linewidth}
		\centering
		\includegraphics[width=0.55in,height=0.55in]{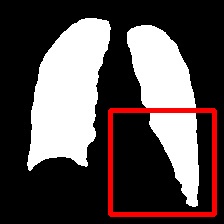}
		\label{MedT5}
	\end{minipage}
	\begin{minipage}{0.11\linewidth}
		\centering
		\includegraphics[width=0.55in,height=0.55in]{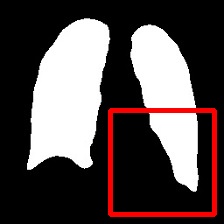}
		\label{GTUNett5}
	\end{minipage}
	\begin{minipage}{0.11\linewidth}
		\centering
		\includegraphics[width=0.55in,height=0.55in]{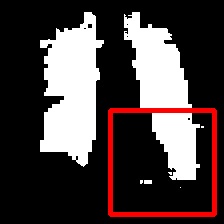}
		\label{SwinUNet5}
	\end{minipage}
	\begin{minipage}{0.11\linewidth}
		\centering
		\includegraphics[width=0.55in,height=0.55in]{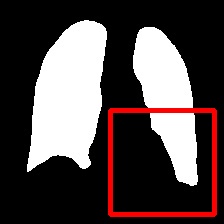}
		\label{UCTransNet5}
	\end{minipage}
	\begin{minipage}{0.11\linewidth}
		\centering
		\includegraphics[width=0.55in,height=0.55in]{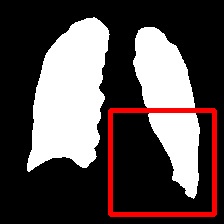}
		\label{LViT5}
	\end{minipage}
	\begin{minipage}{0.11\linewidth}
		\centering
		\includegraphics[width=0.55in,height=0.55in]{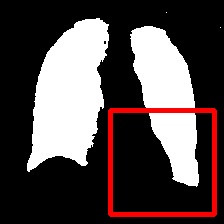}
		\label{OURS5}
	\end{minipage}
	\begin{minipage}{0.11\linewidth}
		\centering
		\includegraphics[width=0.55in,height=0.55in]{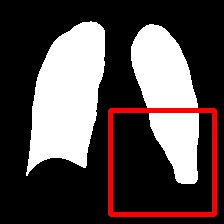}
		\label{GT5}
	\end{minipage}
	
	\vspace{0.1cm}
	\hspace{-0.2cm}
	\begin{minipage}{0.05\linewidth}
		\centering
		\vspace{-0.7cm}
		\centerline{{\scriptsize \makecell{The two\\ infected regions\\ are symmetric.}}}
	\end{minipage}
	\begin{minipage}{0.11\linewidth}
		\centering
		\includegraphics[width=0.55in,height=0.55in]{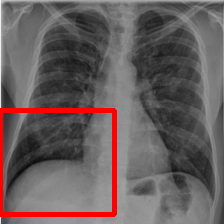}
		\centerline{\scriptsize  IMG}
		\label{IMG6}
	\end{minipage}
	\begin{minipage}{0.11\linewidth}
		\centering
		\includegraphics[width=0.55in,height=0.55in]{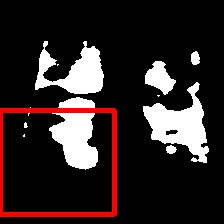}
		\centerline{\scriptsize  MedT}
		\label{MedT6}
	\end{minipage}
	\begin{minipage}{0.11\linewidth}
		\centering
		\includegraphics[width=0.55in,height=0.55in]{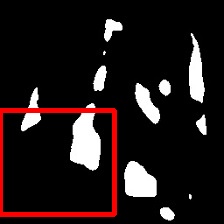}
		\centerline{\scriptsize  GTUNet}
		\label{GTUNet6}
	\end{minipage}
	\begin{minipage}{0.11\linewidth}
		\centering
		\includegraphics[width=0.55in,height=0.55in]{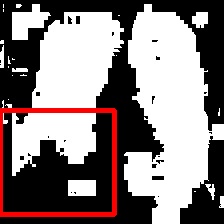}
		\centerline{\scriptsize  SwinUNet}
		\label{SwinUNet6}
	\end{minipage}
	\begin{minipage}{0.11\linewidth}
		\centering
		\includegraphics[width=0.55in,height=0.55in]{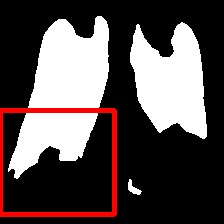}
		\centerline{\scriptsize  UCTransNet}
		\label{UCTransNet6}
	\end{minipage}
	\begin{minipage}{0.11\linewidth}
		\centering
		\includegraphics[width=0.55in,height=0.55in]{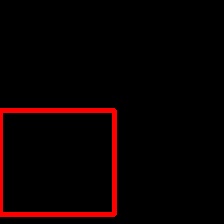}
		\centerline{\scriptsize  LViT}
		\label{LViT6}
	\end{minipage}
	\begin{minipage}{0.11\linewidth}
		\centering
		\includegraphics[width=0.55in,height=0.55in]{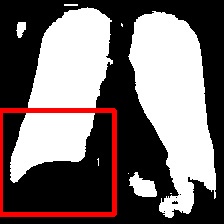}
		\centerline{\scriptsize  OURS}
		\label{OURS6}
	\end{minipage}
	\begin{minipage}{0.11\linewidth}
		\centering
		\includegraphics[width=0.55in,height=0.55in]{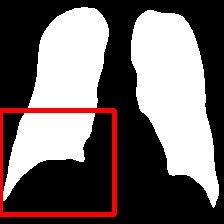}
		\centerline{\scriptsize  GT}
		\label{GT6}
	\end{minipage}
	\caption{Medical image segmentation from QaTa-COV19 dataset. The text annotations are created by hand-craft.}
	\vspace{-0.3cm}
	\label{test2}
\end{figure*}

\subsubsection{UpAttention}
As shown in Fig. \ref{fig.UpAttention}, UpAttention is designed to refine the progressive features for high-quality masks. Based on the convolutional block attention module (CBAM \cite{woo2018cbam}), it contains parallel pieces for global average pooling (GAP) and global max pooling (GMP).
First, $tensor_2$ and $tensor_3$ are sent into the block. $Tensor_3$ is upsampled into $B \times 12C \times  \frac{H}{2} \times \frac{W}{2}$. $Tensor_2$ goes through the GAP and GMP. The two outputs are added together as the sum. Then, we concatenate these three and perform a linear function and ReLU activation. Finally, the output will be multiplied by the original $tensor_2$.
After the operations on $tensor_2$ and $tensor_3$, we concatenate them to pass through ConvBN blocks.
Th results from the UpAttention above and $tensor_1$ are paired to be the inputs of the next UpAttention.
The result of this block is our final segmented mask.

In general, our model contains two stages. $Stage 1$ takes contrastive learning as a reference to pretrain PPE based on bountiful amounts of natural image-text pairs. Hence, PPE has the extraordinary ability to seize practical representations, which helps generate single-scale multimodality features. The CNN branch and U-shaped Transformer branch are combined to constitute the basic architecture of PPE. This design can better encode global features with Transformer while retaining the CNN's ability to extract local features.
In $Stage 2$, PPE inherited from $Stage 1$ uses the medical text as prior prompts to generate single-scale features. Then, MSFF develops them into multiscale multimodality features to exchange information. This progressive operation helps compensate for the semantic gap between the natural data and medical data, and contributes to high-quality masks. The UpAttention block is designed to refine the predicted masks.

\section{Experiments and Result}
\subsection{Dataset details}
We first pretrain our PPE on the COCO dataset \cite{lin2014microsoft}. Then, we conduct our experiments and illustrate the performance of our model on three datasets, i.e., the MoNuSeg (microscopic) \cite{kumar2017dataset, kumar2019multi}, the QaTa-COV19 dataset \cite{degerli2022osegnet} and the Glas dataset \cite{sirinukunwattana2017gland}. To demonstrate that our model is also effective on natural images, we conduct experiments on CoSKEL \cite{jerripothula2017object} and MFFD \cite{dias2018multispecies}.

\textbf{Text annotations:} Since the MoNuSeg dataset has annotations, while other medical image datasets do not, we have adopted the same annotation pattern for other medical images. In addition, the simplicity of the mask in medical images allows for a more focused approach to annotations, with a primary focus on segmentation regions and location information. In contrast, natural image text annotations are typically more complex, so we put emphasis on the concrete objects in pictures instead of position information. Therefore, we have employed BLIP \cite{li2022blip}, a highly popular image-to-text generative model, to generate these annotations. The descriptions of these datasets are presented as follows:

\textbf{Coco:} It collects images of common objects in complex everyday scenes, which includes 91 object types and is easy to distinguish. Each image has five written caption descriptions, in which we choose the first one. Then, $5000$ images are picked up as our training set. We process the downloaded files and extract the delineation of the segmented parts. The descriptions are put into Excel. When reading an image, it is resized into $224\times224$, and its corresponding text statistic is also fetched to guide the segmentation task.

\textbf{MoNuSeg Dataset:} This dataset introduces hematoxylin and eosin ($H\&E$)-stained tissue images with more than 21000 painstakingly annotated nuclear boundaries. We select 24 images from its training set$\left(30\ tissue\ images\right)$ as our own training set. It contains 7 organs, breast, liver, kidney, prostate, bladder, colon and stomach. The 24 images are resized into $224\times224$ before being fed into the model. In terms of the corresponding medical terminologies, they are first split and put into a list. Then, we obtained the text token after the text list was processed by BertEmbedding \cite{devlin2018bert}. The test set has 10 images. When the indices reach the top, the predictions on the test set are saved into the corresponding file.

\textbf{QaTa-COV19:} This dataset from Tampere University and Qatar University contains $121,378$ chest X-ray images (CXRs), including 9258 coronavirus disease 2019 (COVID-19) samples. Each image has a corresponding ground-truth mask for COVID-19 pneumonia segmentation. We choose 50 images as our training set and 10 images as our testing set. In addition, we create the text annotations for this datasets.

\textbf{Glas:} This dataset \cite{sirinukunwattana2017gland} is created to encourage research in gland segmentation algorithms. It comprises 37 benign and 48 malignant training images, with the testing dataset consisting of 37 benign and 43 malignant images. We randomly selected 35 images for our training set and 11 images for the test set. Furthermore, we provided detailed text annotations for each image.

\textbf{CoSKEL:} This dataset contains $26$ biological categories, including people, animals and plants. We randomly select $40$ images, such as a bird, panda, iris, fire pink, and so on. The test set includes $9$ images. To prove the effectiveness of our language-driven model, BLIP \cite{li2022blip} is applied to generate corresponding text annotations.

\textbf{MFFD:} This natural images dataset comes from multispecies fruit flower detection \cite{dias2018multispecies}, and includes $Apple$, $Peach$ and $Pear$. Images of different fruit flower species were collected under different angles of capture. We randomly select $26$ images from it for the training process and $10$ images for the test. Before they are sent to the model, the image size is reshaped into $224\times224$. Text annotations are generated from BLIP \cite{li2022blip}.

\subsection{Loss Function}
\textbf{Weighted Binary Cross Entropy Loss:} We use the weighted binary cross entropy (WBCE) function to measure the discrepancy between the outputs of our model and the corresponding ground truths. Given the input image $x$ and its ground truth vector $y$, the BCE loss formula is as follows:
\begin{equation}\label{BCEloss}
	\begin{aligned}
		L_{BCE(x,y)} = -\frac{1}{N} \sum_{i=0}^{N-1}(y_{i}log(p(x_{i})))+(1-y_{i})log(1-p(x_{i}))
	\end{aligned}
\end{equation}
where $N$ represents the batch number, $p(x_{i})$ is for the predictions of the inputs. 

After computing the BCE loss, we multiply the weight values for different pixel values according to the ground truth. We apply the following formula:
\begin{equation}\label{WBCEloss}
	\begin{aligned}
		L_{WBCE(x,y)} = \frac{w_{1}\cdot pos\cdot L_{BCE(x,y)}}{pos_{sum}} +
		\frac{w_{2}\cdot neg\cdot L_{BCE(x,y)}}{neg_{sum}}
	\end{aligned}
\end{equation}
where $w_{1}$ and $w_{2}$ denote the weights, which are both equal to $0.5$ in this paper. $Pos$ and $neg$ indicate the positive and negative pixel values of the ground truth. $Pos_{sum}$ and $neg_{sum}$ are the sum results of $pos$ and $neg$. This function works well because segmentation is a pixel-level classification.

\begin{figure*}[h!]
	\centering
	\begin{minipage}{0.05\linewidth}
		\centering
		\centerline{{\scriptsize \makecell{The nuclei are\\ irregularly\\ distributed.}}}
	\end{minipage}
	\begin{minipage}{0.11\linewidth}
		\centering
		\includegraphics[width=0.55in,height=0.55in]{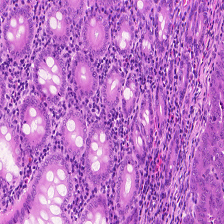}
		\label{IMG7}
	\end{minipage}
	\begin{minipage}{0.11\linewidth}
		\centering
		\includegraphics[width=0.55in,height=0.55in]{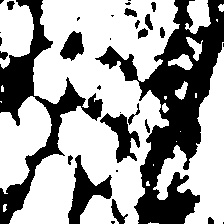}
		\label{MedT7}
	\end{minipage}
	\begin{minipage}{0.11\linewidth}
		\centering
		\includegraphics[width=0.55in,height=0.55in]{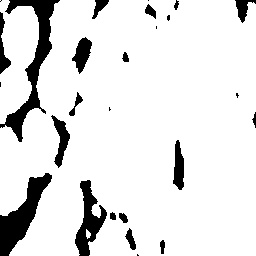}
		\label{GTUNet7}
	\end{minipage}
	\begin{minipage}{0.11\linewidth}
		\centering
		\includegraphics[width=0.55in,height=0.55in]{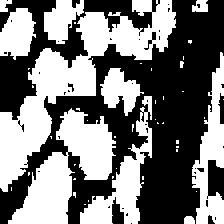}
		\label{SwinUNet7}
	\end{minipage}
	\begin{minipage}{0.11\linewidth}
		\centering
		\includegraphics[width=0.55in,height=0.55in]{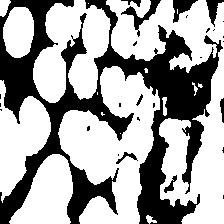}
		\label{UCTransNet7}
	\end{minipage}
	\begin{minipage}{0.11\linewidth}
		\centering
		\includegraphics[width=0.55in,height=0.55in]{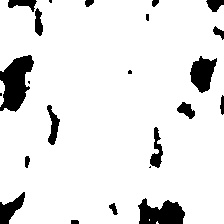}
		\label{LViT7}
	\end{minipage}
	\begin{minipage}{0.11\linewidth}
		\centering
		\includegraphics[width=0.55in,height=0.55in]{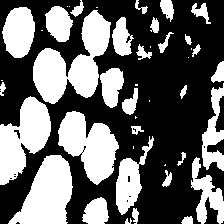}
		\label{OURS7}
	\end{minipage}
	\begin{minipage}{0.11\linewidth}
		\centering
		\includegraphics[width=0.55in,height=0.55in]{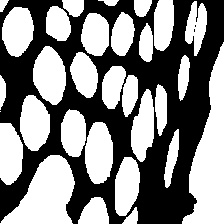}
		\label{GT7}
	\end{minipage}
	
	\vspace{0.1cm}
	\begin{minipage}{0.05\linewidth}
		\centering
		\centerline{{\scriptsize \makecell{The nuclei are\\ evenly\\ distributed.}}}
	\end{minipage}
	\begin{minipage}{0.11\linewidth}
		\centering
		\includegraphics[width=0.55in,height=0.55in]{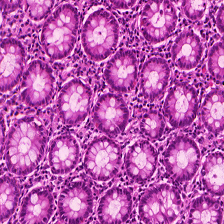}
		\label{IMG8}
	\end{minipage}
	\begin{minipage}{0.11\linewidth}
		\centering
		\includegraphics[width=0.55in,height=0.55in]{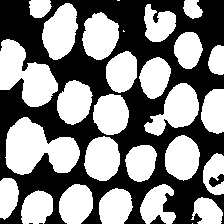}
		\label{MedT8}
	\end{minipage}
	\begin{minipage}{0.11\linewidth}
		\centering
		\includegraphics[width=0.55in,height=0.55in]{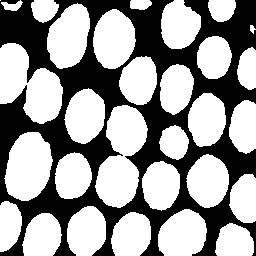}
		\label{GTUNet8}
	\end{minipage}
	\begin{minipage}{0.11\linewidth}
		\centering
		\includegraphics[width=0.55in,height=0.55in]{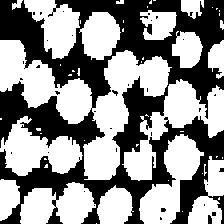}
		\label{SwinUNet8}
	\end{minipage}
	\begin{minipage}{0.11\linewidth}
		\centering
		\includegraphics[width=0.55in,height=0.55in]{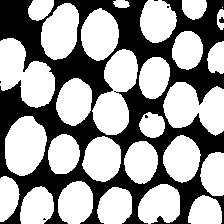}
		\label{UCTransNet8}
	\end{minipage}
	\begin{minipage}{0.11\linewidth}
		\centering
		\includegraphics[width=0.55in,height=0.55in]{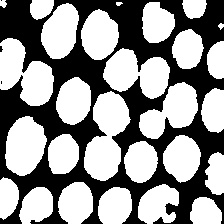}
		\label{LViT8}
	\end{minipage}
	\begin{minipage}{0.11\linewidth}
		\centering
		\includegraphics[width=0.55in,height=0.55in]{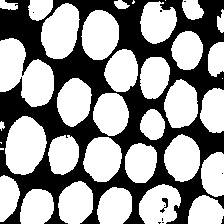}
		\label{OURS8}
	\end{minipage}
	\begin{minipage}{0.11\linewidth}
		\centering
		\includegraphics[width=0.55in,height=0.55in]{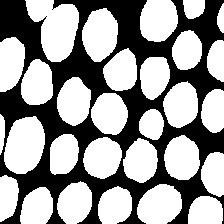}
		\label{GT8}
	\end{minipage}
	
	\vspace{0.1cm}
	\begin{minipage}{0.05\linewidth}
		\centering
		\centerline{{\scriptsize \makecell{The nuclei are\\ sparsely\\ distributed.}}}
	\end{minipage}
	\begin{minipage}{0.11\linewidth}
		\centering
		\includegraphics[width=0.55in,height=0.55in]{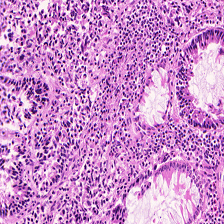}
		\label{IMG9}
	\end{minipage}
	\begin{minipage}{0.11\linewidth}
		\centering
		\includegraphics[width=0.55in,height=0.55in]{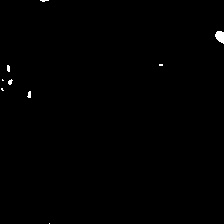}
		\label{MedT9}
	\end{minipage}
	\begin{minipage}{0.11\linewidth}
		\centering
		\includegraphics[width=0.55in,height=0.55in]{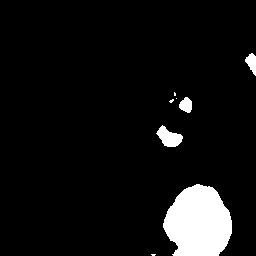}
		\label{GTUNet9}
	\end{minipage}
	\begin{minipage}{0.11\linewidth}
		\centering
		\includegraphics[width=0.55in,height=0.55in]{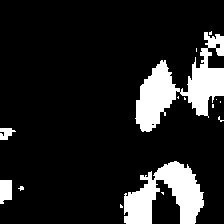}
		\label{SwinUNet9}
	\end{minipage}
	\begin{minipage}{0.11\linewidth}
		\centering
		\includegraphics[width=0.55in,height=0.55in]{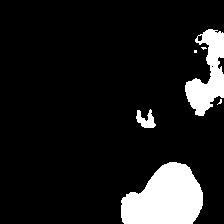}
		\label{UCTransNet9}
	\end{minipage}
	\begin{minipage}{0.11\linewidth}
		\centering
		\includegraphics[width=0.55in,height=0.55in]{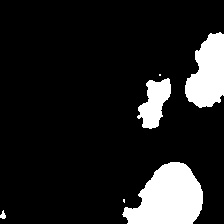}
		\label{LViT9}
	\end{minipage}
	\begin{minipage}{0.11\linewidth}
		\centering
		\includegraphics[width=0.55in,height=0.55in]{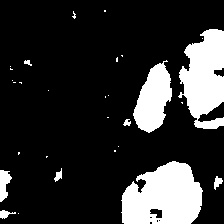}
		\label{OURS9}
	\end{minipage}
	\begin{minipage}{0.11\linewidth}
		\centering
		\includegraphics[width=0.55in,height=0.55in]{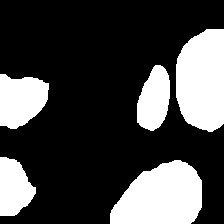}
		\label{GT9}
	\end{minipage}
	
	\vspace{0.1cm}
	\hspace{-0.2cm}
	\begin{minipage}{0.05\linewidth}
		\centering
		\vspace{-0.7cm}
		\centerline{{\scriptsize \makecell{The nuclei are\\ irregularly\\ distributed.}}}
	\end{minipage}
	\begin{minipage}{0.11\linewidth}
		\centering
		\includegraphics[width=0.55in,height=0.55in]{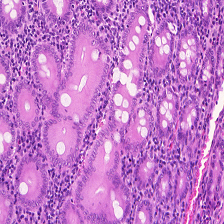}
		\centerline{\scriptsize  IMG}
		\label{IMG10}
	\end{minipage}
	\begin{minipage}{0.11\linewidth}
		\centering
		\includegraphics[width=0.55in,height=0.55in]{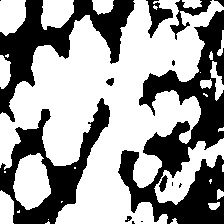}
		\centerline{\scriptsize  MedT}
		\label{MedT10}
	\end{minipage}
	\begin{minipage}{0.11\linewidth}
		\centering
		\includegraphics[width=0.55in,height=0.55in]{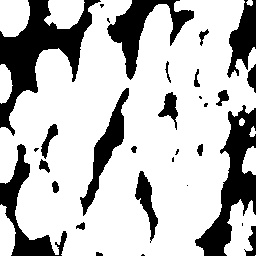}
		\centerline{\scriptsize  GTUNet}
		\label{GTUNet10}
	\end{minipage}
	\begin{minipage}{0.11\linewidth}
		\centering
		\includegraphics[width=0.55in,height=0.55in]{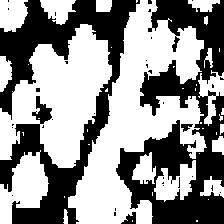}
		\centerline{\scriptsize  SwinUNet}
		\label{SwinUNet10}
	\end{minipage}
	\begin{minipage}{0.11\linewidth}
		\centering
		\includegraphics[width=0.55in,height=0.55in]{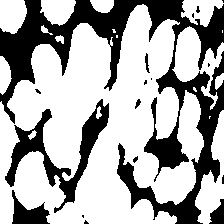}
		\centerline{\scriptsize  UCTransNet}
		\label{UCTransNet10}
	\end{minipage}
	\begin{minipage}{0.11\linewidth}
		\centering
		\includegraphics[width=0.55in,height=0.55in]{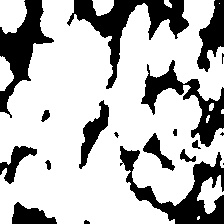}
		\centerline{\scriptsize  LViT}
		\label{LViT10}
	\end{minipage}
	\begin{minipage}{0.11\linewidth}
		\centering
		\includegraphics[width=0.55in,height=0.55in]{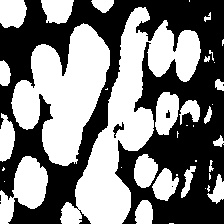}
		\centerline{\scriptsize  OURS}
		\label{OURS10}
	\end{minipage}
	\begin{minipage}{0.11\linewidth}
		\centering
		\includegraphics[width=0.55in,height=0.55in]{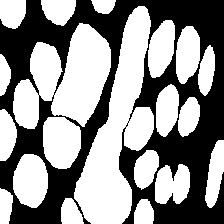}
		\centerline{\scriptsize  GT}
		\label{GT10}
	\end{minipage}
	%\vspace{-0.3cm}
	\caption{Medical image segmentation from Glas dataset. The text annotations are created by hand-craft.}
	\vspace{-0.3cm}
	\label{test3}
\end{figure*}

\textbf{Weighted Dice Loss:} The Dice coefficient is a measurement function used to evaluate the similarity of two samples. Dice loss is calculated from it and can present a good performance when the positive and negative samples are imbalanced. Given the input image $x$ and its ground truth vector $y$, the weighted Dice loss formula is as follows:
\begin{equation}\label{WDloss}
	\begin{aligned}
		L_{WDiCE(x,y)} = 1 -\frac{1}{N} \sum_{i=0}^{N-1}\frac{w_{1}\ p(x_{i}) \cdot w_{2}\cdot y_{i} + smooth}{w_{1} \cdot p(x_{i}^{2}) + w_{2}\cdot y_{i}^{2} + smooth}
	\end{aligned}
\end{equation}
where $w_{1}$ and $w_{2}$ are also set to $0.5$. $Smooth$ is a Laplace smoothing parameter, which is equal to $10^{-12}$ for a zero probability.

After calculating the two losses, we obtain the final loss value:
\begin{equation}\label{WDloss}
	\begin{aligned}
		L = w_{1} \cdot L_{WBCE(x,y)} + w_{2} \cdot L_{WDiCE(x,y)}
	\end{aligned}
\end{equation}
where the weights are both $0.5$.

\subsection{Evaluation Metrics}
We use accuracy (Acc), mean intersection over union (mIoU) and Dice values to measure our experiments. The metric formulations are as follows:
\begin{equation}\label{acc}
	\centering
	Acc = \frac{TP+TN}{TP+FN+FP+TN}
\end{equation} 
\begin{equation}\label{mIoU}
	\centering
	\begin{aligned}
		mIoU = \frac{1}{2} \Big\{\frac{TP}{TP+FP+FN}+\frac{TN}{TN+FN+FP}\Big\}
	\end{aligned}
\end{equation}
\begin{equation}\label{dice}
	\centering
	dice = 2 \cdot \frac{X \cdot Y}{X + Y}
\end{equation} 
where TP, FN, FP and TN denote the number of true positives, false-negatives, false-positives and true negatives, respectively. $X$ and $Y$ denote the input images and the output predictions, respectively.

\begin{figure*}[h!]
	\centering
	\begin{minipage}{0.05\linewidth}
		\centering
		\centerline{{\scriptsize \makecell{A girl\\ playing\\ soccer.}}}
	\end{minipage}
	\begin{minipage}{0.11\linewidth}
		\centering
		\includegraphics[width=0.55in,height=0.55in]{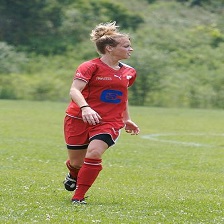}
	\end{minipage}
	\begin{minipage}{0.11\linewidth}
		\centering
		\includegraphics[width=0.55in,height=0.55in]{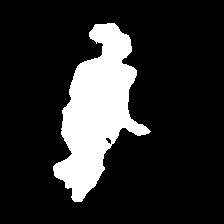}
	\end{minipage}
	\begin{minipage}{0.11\linewidth}
		\centering
		\includegraphics[width=0.55in,height=0.55in]{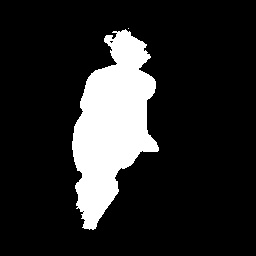}
	\end{minipage}
	\begin{minipage}{0.11\linewidth}
		\centering
		\includegraphics[width=0.55in,height=0.55in]{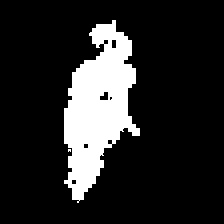}
	\end{minipage}
	\begin{minipage}{0.11\linewidth}
		\centering
		\includegraphics[width=0.55in,height=0.55in]{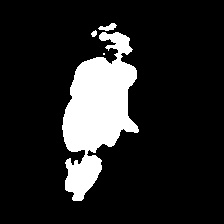}
	\end{minipage}
	\begin{minipage}{0.11\linewidth}
		\centering
		\includegraphics[width=0.55in,height=0.55in]{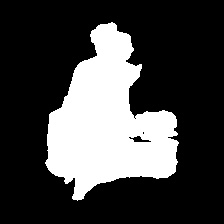}
	\end{minipage}
	\begin{minipage}{0.11\linewidth}
		\centering
		\includegraphics[width=0.55in,height=0.55in]{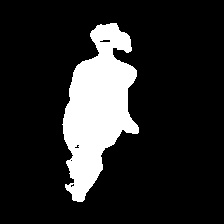}
	\end{minipage}
	\begin{minipage}{0.11\linewidth}
		\centering
		\includegraphics[width=0.55in,height=0.55in]{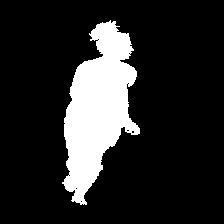}
	\end{minipage}
	
	\vspace{0.1cm}
	\begin{minipage}{0.05\linewidth}
		\centering
		\centerline{{\scriptsize \makecell{A red flower\\ in the woods.}}}
	\end{minipage}
	\begin{minipage}{0.11\linewidth}
		\centering
		\includegraphics[width=0.55in,height=0.55in]{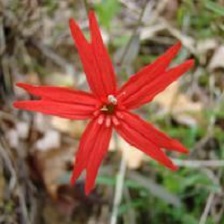}
	\end{minipage}
	\begin{minipage}{0.11\linewidth}
		\centering
		\includegraphics[width=0.55in,height=0.55in]{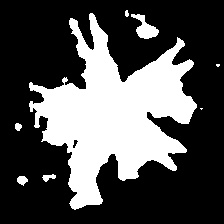}
	\end{minipage}
	\begin{minipage}{0.11\linewidth}
		\centering
		\includegraphics[width=0.55in,height=0.55in]{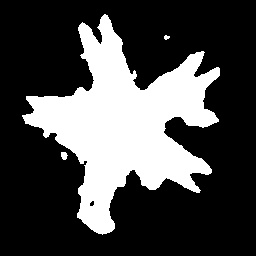}
	\end{minipage}
	\begin{minipage}{0.11\linewidth}
		\centering
		\includegraphics[width=0.55in,height=0.55in]{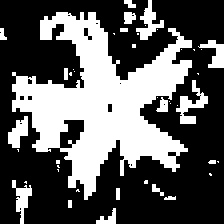}
	\end{minipage}
	\begin{minipage}{0.11\linewidth}
		\centering
		\includegraphics[width=0.55in,height=0.55in]{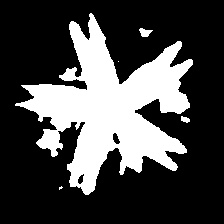}
	\end{minipage}
	\begin{minipage}{0.11\linewidth}
		\centering
		\includegraphics[width=0.55in,height=0.55in]{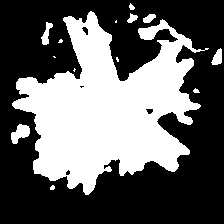}
	\end{minipage}
	\begin{minipage}{0.11\linewidth}
		\centering
		\includegraphics[width=0.55in,height=0.55in]{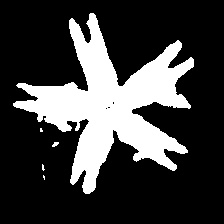}
	\end{minipage}
	\begin{minipage}{0.11\linewidth}
		\centering
		\includegraphics[width=0.55in,height=0.55in]{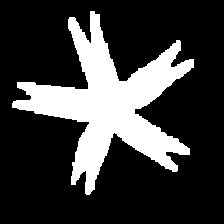}
	\end{minipage}
	
	\vspace{0.1cm}
	%\hspace{-0.25cm}
	\begin{minipage}{0.05\linewidth}
		\centering
		\vspace{-0.3cm}
		\centerline{{\scriptsize \makecell{A cat is\\ walking\\ on the street.}}}
	\end{minipage}
	\begin{minipage}{0.11\linewidth}
		\centering
		\includegraphics[width=0.55in,height=0.55in]{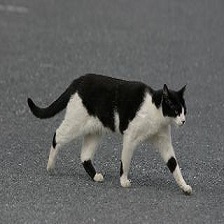}
		\centerline{\scriptsize  IMG}
	\end{minipage}
	\begin{minipage}{0.11\linewidth}
		\centering
		\includegraphics[width=0.55in,height=0.55in]{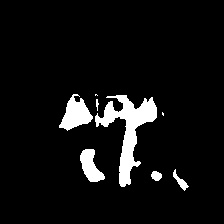}
		\centerline{\scriptsize  MedT}
	\end{minipage}
	\begin{minipage}{0.11\linewidth}
		\centering
		\includegraphics[width=0.55in,height=0.55in]{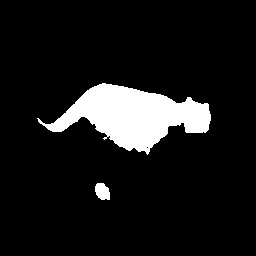}
		\centerline{\scriptsize  GTUNet}
	\end{minipage}
	\begin{minipage}{0.11\linewidth}
		\centering
		\includegraphics[width=0.55in,height=0.55in]{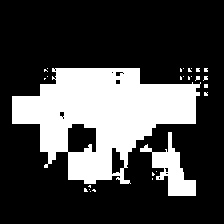}
		\centerline{\scriptsize  SwinUNet}
	\end{minipage}
	\begin{minipage}{0.11\linewidth}
		\centering
		\includegraphics[width=0.55in,height=0.55in]{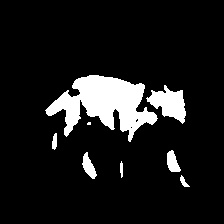}
		\centerline{\scriptsize  UCTransNet}
	\end{minipage}
	\begin{minipage}{0.11\linewidth}
		\centering
		\includegraphics[width=0.55in,height=0.55in]{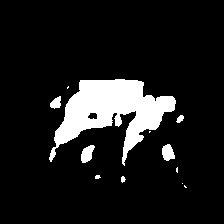}
		\centerline{\scriptsize  LViT}
	\end{minipage}
	\begin{minipage}{0.11\linewidth}
		\centering
		\includegraphics[width=0.55in,height=0.55in]{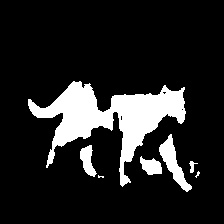}
		\centerline{\scriptsize  OURS}
	\end{minipage}
	\begin{minipage}{0.11\linewidth}
		\centering
		\includegraphics[width=0.55in,height=0.55in]{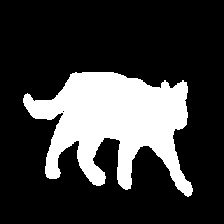}
		\centerline{\scriptsize  GT}
	\end{minipage}
	\caption{Natural image segmentation from CoSKEL dataset. BLIP \cite{li2022blip} is applied to generate corresponding text annotations.}
	\label{test4}
\end{figure*}
\begin{figure*}[h!]
	\centering
	\begin{minipage}{0.05\linewidth}
		\centering
		\centerline{{\scriptsize \makecell{The flowers are\\ irregularly\\ distributed.}}}
	\end{minipage}
	\begin{minipage}{0.11\linewidth}
		\centering
		\includegraphics[width=0.55in,height=0.55in]{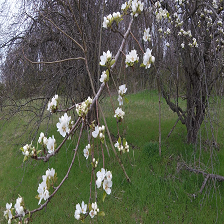}
	\end{minipage}
	\begin{minipage}{0.11\linewidth}
		\centering
		\includegraphics[width=0.55in,height=0.55in]{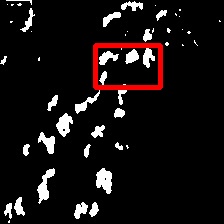}
	\end{minipage}
	\begin{minipage}{0.11\linewidth}
		\centering
		\includegraphics[width=0.55in,height=0.55in]{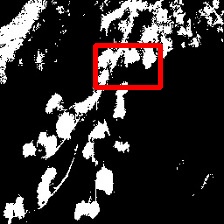}
	\end{minipage}
	\begin{minipage}{0.11\linewidth}
		\centering
		\includegraphics[width=0.55in,height=0.55in]{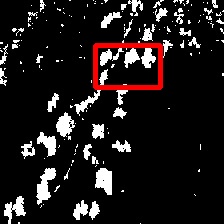}
	\end{minipage}
	\begin{minipage}{0.11\linewidth}
		\centering
		\includegraphics[width=0.55in,height=0.55in]{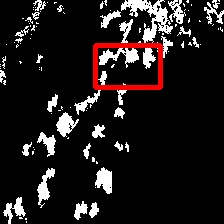}
	\end{minipage}
	\begin{minipage}{0.11\linewidth}
		\centering
		\includegraphics[width=0.55in,height=0.55in]{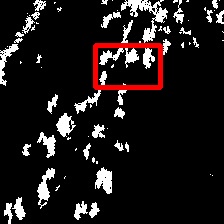}
	\end{minipage}
	\begin{minipage}{0.11\linewidth}
		\centering
		\includegraphics[width=0.55in,height=0.55in]{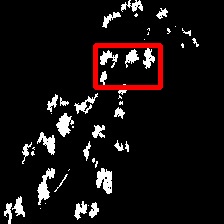}
	\end{minipage}
	\begin{minipage}{0.11\linewidth}
		\centering
		\includegraphics[width=0.55in,height=0.55in]{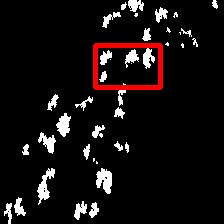}
	\end{minipage}
	
	\vspace{0.1cm}
	\begin{minipage}{0.05\linewidth}
		\centering
		\centerline{{\scriptsize \makecell{The flowers are\\ sparsely\\ distributed.}}}
	\end{minipage}
	\begin{minipage}{0.11\linewidth}
		\centering
		\includegraphics[width=0.55in,height=0.55in]{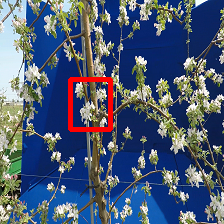}
	\end{minipage}
	\begin{minipage}{0.11\linewidth}
		\centering
		\includegraphics[width=0.55in,height=0.55in]{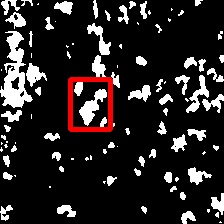}
	\end{minipage}
	\begin{minipage}{0.11\linewidth}
		\centering
		\includegraphics[width=0.55in,height=0.55in]{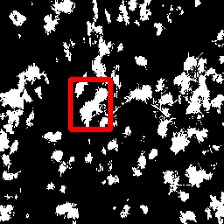}
	\end{minipage}
	\begin{minipage}{0.11\linewidth}
		\centering
		\includegraphics[width=0.55in,height=0.55in]{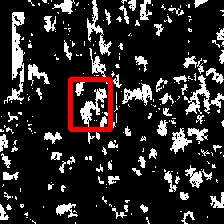}
	\end{minipage}
	\begin{minipage}{0.11\linewidth}
		\centering
		\includegraphics[width=0.55in,height=0.55in]{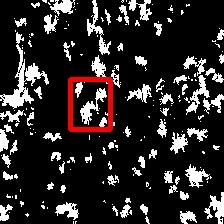}
	\end{minipage}
	\begin{minipage}{0.11\linewidth}
		\centering
		\includegraphics[width=0.55in,height=0.55in]{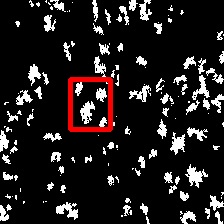}
	\end{minipage}
	\begin{minipage}{0.11\linewidth}
		\centering
		\includegraphics[width=0.55in,height=0.55in]{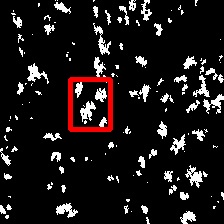}
	\end{minipage}
	\begin{minipage}{0.11\linewidth}
		\centering
		\includegraphics[width=0.55in,height=0.55in]{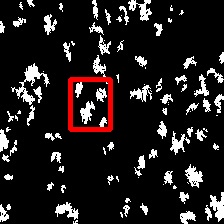}
	\end{minipage}
	
	\vspace{0.1cm}
	\begin{minipage}{0.05\linewidth}
		\centering
		\vspace{-0.3cm}
		\centerline{{\scriptsize \makecell{The flower\\ density in the \\upper part is high.}}}
	\end{minipage}
	\begin{minipage}{0.11\linewidth}
		\centering
		\includegraphics[width=0.55in,height=0.55in]{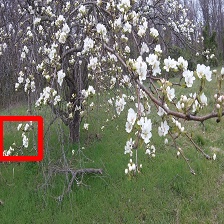}
		\centerline{\scriptsize  IMG}
	\end{minipage}
	\begin{minipage}{0.11\linewidth}
		\centering
		\includegraphics[width=0.55in,height=0.55in]{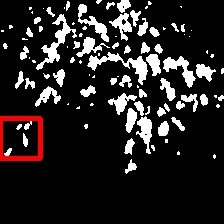}
		\centerline{\scriptsize  MedT}
	\end{minipage}
	\begin{minipage}{0.11\linewidth}
		\centering
		\includegraphics[width=0.55in,height=0.55in]{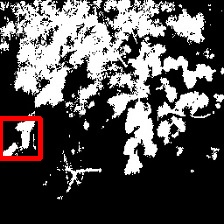}
		\centerline{\scriptsize  GTUNet}
	\end{minipage}
	\begin{minipage}{0.11\linewidth}
		\centering
		\includegraphics[width=0.55in,height=0.55in]{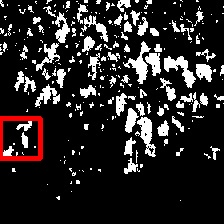}
		\centerline{\scriptsize  SwinUNet}
	\end{minipage}
	\begin{minipage}{0.11\linewidth}
		\centering
		\includegraphics[width=0.55in,height=0.55in]{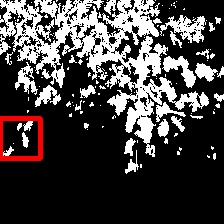}
		\centerline{\scriptsize  UCTransNet}
	\end{minipage}
	\begin{minipage}{0.11\linewidth}
		\centering
		\includegraphics[width=0.55in,height=0.55in]{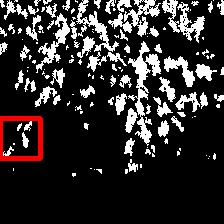}
		\centerline{\scriptsize  LViT}
	\end{minipage}
	\begin{minipage}{0.11\linewidth}
		\centering
		\includegraphics[width=0.55in,height=0.55in]{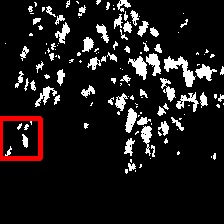}
		\centerline{\scriptsize  OURS}
	\end{minipage}
	\begin{minipage}{0.11\linewidth}
		\centering
		\includegraphics[width=0.55in,height=0.55in]{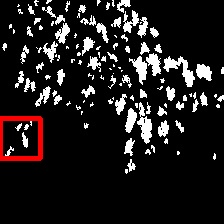}
		\centerline{\scriptsize  GT}
	\end{minipage}
	\caption{Natural image segmentation from MFFD dataset. We use BLIP \cite{li2022blip} to produce text annotations.}
	\label{test5}
\end{figure*}

\subsection{Qualitative Results}
We randomly select some results in the comparison experiments. Fig. \ref{test1}, Fig. \ref{test2}, and Fig. \ref{test3} show that our model presents the best visual effect on the MoNuSeg, QaTa-COV19 and Glas datasets. To validate the efficiency on natural images, we also conduct experiments on the CoSKEL and MFFD datasets in Fig. \ref{test4} and Fig. \ref{test5}.

As illustrated in Fig. \ref{test1}, MedT \cite{valanarasu2021medical} is an attention-based model that adds gated parameters to control the self-attention module. This design helps model the long-range dependencies, so MedT behaves well in capturing the global features. Nevertheless, it cannot remedy the defect of lacking local statistics. Therefore, some minor nuclear boundaries may be neglected, and some white pieces emerge from the final maps.

GTUNet \cite{li2021gt} uses group Transformer to reduce the computational cost of the traditional Transformer architectures, for which the implementation merits both Transformer and UNet. However, as shown in Fig. \ref{test1}, there are some tiny mistakes, which are some extra segmented areas. This is because GTUNet emphasizes global features, leading to unsatisfactory results.

Similar to GTUNet, SwinUNet \cite{cao2021swin} also leverages Transformer to analyze global features but is a pure Transformer-based model. It treats the Swin Transformer as a basic block to construct a U-shape network. The encoder in the architecture extracts multiscale representations. Then, the decoder with the patch embedding layer fuses them to restore the spatial resolution of the feature maps. Therefore, SwinUNet has a better visual effect compared with the other methods. Nevertheless, it has some trouble in depicting edge information, and presents indistinct shapes.

UCTransNet \cite{wang2022uctransnet} ameliorates not only UNet but also Transformer. Wang \emph{et al.} \cite{wang2022uctransnet} deemed that not every skip connection set in UNet is effective. For this reason, they designed a channel transformer (CTrans) to substitute for the skip connections. The results are more accurate than those of the first three methods in some examples. However, UCTransNet struggles to segment some high contrast ratio images.

LViT \cite{li2022lvit} introduces language-image pair input to guide the segmentation task. Medical text annotations are employed to compensate for the inadequacies present in image data due to the high accuracy requirements of medical image segmentation. As a result, LViT is able to capture certain details. However, the scarcity of medical data leads to disappointing outcomes. 

Our language-driven model incorporates the benefits of text prior prompts and leverages medical terminology to improve segmentation accuracy while also controlling computational costs. Additionally, we utilize CL to pretrain PPE on extensive natural image-text pairs, which can extract valuable representations after the model has converged. In the downstream stage, PPE inherits the network parameters of the pretrained PPE to process medical data. Thanks to the CL and pretraining phases, our model is capable of achieving the most accurate edge delineation.

\begin{table*}[!h]
	\centering
	\renewcommand{\arraystretch}{1.5}
	\caption{Quantitative comparison of baselines and our model on medical datasets}
	\resizebox{\linewidth}{!}{
		\begin{tabular}{c|ccccc|ccccc|ccccc}
			\hline
			\textbf{Network} & \multicolumn{5}{c|}{\textbf{MoNuSeg}}&\multicolumn{5}{c|}{\textbf{QaTa-COV19}}&\multicolumn{5}{c}{\textbf{Glas}}\\ \cline{1-16}  
			& \textbf{Dice(\%)} & \textbf{mIoU(\%)} & \textbf{Acc(\%)} & \textbf{Precision(\%)} & \textbf{Recall(\%)} 
			& \textbf{Dice(\%)} & \textbf{mIoU(\%)} & \textbf{Acc(\%)} & \textbf{Precision(\%)} & \textbf{Recall(\%)}          
			& \textbf{Dice(\%)} & \textbf{mIoU(\%)} & \textbf{Acc(\%)} & \textbf{Precision(\%)} & \textbf{Recall(\%)}\\
			MedT \cite{valanarasu2021medical} & 72.06  & 56.51  & 85.26  & 95.57  & 90.62
			& 84.90  & 76.05  & 90.21  & 97.36  & 97.30
			& 79.47  & 70.87  & 86.35  & 93.54  & 93.21\\
			GTUNet \cite{li2021gt} & 77.26 & 63.14 & 89.24 & 95.37 & 91.76
			& 84.16 & 77.65 & 90.77 & 93.07 & 95.49
			& 83.90 & 75.23 & 86.80 & 93.14 & 94.09\\
			SwinUNet \cite{cao2021swin} 
			& 79.33 & 65.89 & 90.58 & 94.90 & 92.81
			& 82.21 & 70.25 & 87.18 & 93.23 & 92.30
			& 85.77 & 75.99 & 88.25 & 91.82 & 92.23\\
			UCTranNet \cite{wang2022uctransnet} & 73.44 & 58.32 & 85.08 & 96.52 & 93.32
			& 91.49 & 85.10 & 94.18 & 97.22 & 97.23
			& 87.72 & 79.41 & 90.00 & 93.30 & 94.05\\
			LViT \cite{li2022lvit} & 68.71 & 52.77 & 81.18 & 94.93 & 92.53
			& 79.04 & 72.38 & 89.02 & 92.82 & 95.22
			& 86.86 & 78.06 & 87.74 & 93.43 & 94.40 \\ \hline
			OURS & \textbf{80.59} & \textbf{67.59} & \textbf{90.85} & \textbf{97.89} & \textbf{94.10}
			& \textbf{91.53} & \textbf{85.67} & \textbf{95.01} & \textbf{98.24} & \textbf{97.34}
			& \textbf{88.12} & \textbf{79.30} & \textbf{89.80} &\textbf{95.56}  & \textbf{95.56}\\ \hline          
		\end{tabular}
	}
	\label{Quantitative result}
\end{table*}

Fig. \ref{test2} shows the results obtained from experiments on the QaTa-COV19 datasets. Closer inspection of the figure shows that MedT has superiority in depicting long-range information. This benefits from the gated parameters. They can control the amount of information that the positional embedding supplies to key, query and value. It is noteworthy that its ability to learn the local features can still be enhanced.

GTUNet still has trouble capturing local features and depicting the exact edge shape. Inferior to GTUNet, SwinUNet can only produce rough infected regions. However, as shown in Fig. \ref{test2}, it struggles to catch local details, so some incorrect areas are delineated, and the edge depiction is coarse.

UCTransNet and LViT increase their segmented accuracy. We can observe that they depict the correct infected regions of the first and second IMGs in Fig. \ref{test2}. What stands out in their result pictures is that they cannot give precise edge information.

Compared with other networks, our model integrates Transformer and CNN into PPE, which ensures that it can obtain meaningful representations. After it has these powerful abilities, we combine PPE and the downstream segmentation tasks. Owing to the pretrained phase and the well-designed decoder, our model presents the most exact segmented results.

Fig. \ref{test3} presents the experiments on the Glas datasets. The methods on the second IMG have good visual effects, however, the results of MedT and SwinUNet have some hollows. In particular, the resolution of SwinUNet is suboptimal. GTUNet attains exemplary segmented results, but some spare parts exist. LViT and our model surpass the other architectures. In terms of other IMGs, it is obvious that our model possesses the most accurate segmentation cell outline.

Fig. \ref{test4} and Fig. \ref{test5} are the results on the CoSKEL and MFFD datasets.  In the CoSKEL dataset, SwinUNet is relatively weak in processing the edge information. LViT also cannot segment the correct target shape. MedT presents a good performance on the first IMG owing to the rearranged self-attention mechanism.
In the MFFD dataset, the mask results of LViT are relatively reasonable. MedT and SwinUNet obviously give extra parts in the upper left. GTUNet and UCTransNet both have mediocre performances on the two datasets.
Our model presents the best segmentation results on MFFD and possesses the most accurate segmentation mask outline compared with the other models on CoSKEL datasets. 

\subsection{Quantitative Comparison Results}
For a fair comparison, we use the recommended parameters in the comparison methods to test the segmentation effects. In addition, we directly evaluate our model on the test dataset and randomly select images in MoNuSeg, QaTa-COV19, Glas and two natural image datasets. 
A closer inspection of the data in Tab. \ref{Quantitative result} and Tab. \ref{Natural result} reveals that our model outperforms the other methods on Dice, mIoU and Acc. Tab. \ref{Natural result} shows experiments on natural images.

\textbf{Dice:} The Dice index is used to calculate the similarity between the samples. MedT benefits from the improved gated axial attention layers as its main building block, which introduces four gates to control the amount of information. GTUNet surpasses MedT for the well-designed grouping structure and the bottleneck structure. SwinUNet obtains an ideal Dice index, which uses symmetric encoders to take a sample of the image features; nevertheless, it is computationally demanding. UCTransNet reconciles the skip connections in UNet and improves the mechanism. It uses Channel Transformer to repair the semantic gap in UNet, which helps UCTransNet obtain $73.44\%$. 

\begin{table}[!h]
	\centering
	\renewcommand{\arraystretch}{1.5}
	\caption{Quantitative comparison of the baselines and our model on natural datasets}
	\vspace{-0.2cm}
	\resizebox{\linewidth}{!}{
		\begin{tabular}{c|ccccc|ccccc}
			\hline
			\textbf{Network} & \multicolumn{5}{c|}{\textbf{CoSKEL}}&\multicolumn{5}{c}{\textbf{MFFD}}\\ \cline{1-11}  
			& \textbf{Dice(\%)} & \textbf{mIoU(\%)} & \textbf{Acc(\%)} & \textbf{Precision(\%)} & \textbf{Recall(\%)} 
			& \textbf{Dice(\%)} & \textbf{mIoU(\%)} & \textbf{Acc(\%)} & \textbf{Precision(\%)} & \textbf{Recall(\%)}\\
			MedT \cite{valanarasu2021medical} & 43.49  & 34.98  & 82.79  & 95.07 & 95.49
			& 69.25  & 53.71  & 94.61  & 98.31 & 95.77\\
			GTUNet \cite{li2021gt} & 60.91 & 49.52 & 85.87 & 97.70 & 96.40
			& 56.34 & 40.01 & 88.20  & 99.63 & 96.61\\
			SwinUNet \cite{cao2021swin} & 69.64 & 57.99 & 83.51 & 97.11 & 92.62
			& 65.67 & 49.31 & 93.52 & 98.69 & 94.42\\
			UCTranNet \cite{wang2022uctransnet} & 58.23 & 48.04 & 81.12 & 96.24 & 97.09
			& 67.11 & 51.19 & 92.91  & 99.47 & 95.60\\
			LViT \cite{li2022lvit} & 54.86 & 43.47 & 80.28 & 96.10 & 95.47
			& 74.40 & 60.09 & 95.32 & 99.13 & 96.53\\ \hline
			OURS & \textbf{79.32} & \textbf{69.36} & \textbf{89.50} & \textbf{98.61} & \textbf{98.34}
			& \textbf{79.92} & \textbf{67.14} & \textbf{97.11} & \textbf{99.77} & \textbf{98.77}\\ \hline          
		\end{tabular}
	}
	\vspace{-0.4cm}
	\label{Natural result}
\end{table}

LViT also uses multimodal medical data to guide the segmentation task but it may not fit in the end-to-end model. Our model trains PPE on copious natural images to obtain single-scale multimodality features. The PPE in the downstream inherits its network parameters, which is effective in presenting high-accuracy results.

In terms of results on natural datasets, SwinUNet obtains the highest value in comparison methods on CoSKEL, and LViT does the same on MFFD. We can observe from Tab. \ref{Natural result} that our model improves the Dice value by a large margin compared with these two networks. 

\textbf{mIoU:} Mean Intersection over Union (mIoU) calculates the mean of the ratios of the intersection and union of all the categories. It is widely used in image segmentation to evaluate model performance. As illustrated in Tab. \ref{Quantitative result}, MedT and UCTransUNet have similar mIoU scores on the MoNuSeg dataset. They both apply an attention mechanism to construct long-range dependencies so that the models can easily catch global statistics. However, they struggle to capture enough features from a single input modality. LViT introduces multimodal data to guide segmentation but is relatively mediocre on the MoNuSeg dataset because LViT lacks the massive image-text pairs to ensure convergence. SwinUNet presents the highest score on the MoNuSeg dataset. It uses the Swin Transformer as the basic block to downscale the input images, and then the extracted features are fused to restore to the original size.

In terms of the QaTa-COV19 dataset, SwinUNet stacks Transformer to emphasize the global information, which may lead to a lack of edge delineation. MedT rearranges the attention mechanism and delves the deeper relations between the patches and the entire image, which helps construct the local and nonlocal dependencies. This enables it to be better to SwinUNet. UCTransNet achieves the highest scores on three indices in the comparison methods. It ameliorates the potential weakness of the skip connections in UNet. This improvement boosts the model performance and obtains $85.10\%$. 

In the Glas dataset, MedT's mIoU is weaker than that of the other networks in the comparison methods. UCTransNet surpasses GTUNet, SwinUNet and LViT because it bridges the semantic and resolution gap between the low-level and high-level features. Owing to reconciling the medical images and terminologies information, LViT scores higher than SwinUNet.

As can be seen from Tab. \ref{Quantitative result}, our model outperforms the other methods. PPE has a strong ability to extract useful representations. Then, its network parameters are employed in the downstream segmentation task. In addition, fusing multimodal inputs is also instrumental in generating satisfactory values.

\textbf{Acc:} Accuracy (Acc) is the evaluation standard for predicting the accuracy of pixels because segmentation is a pixel-level task. Its calculation comprises both false positives and false negatives, which are commonly used to evaluate the effectiveness of models and algorithms. SwinUNet has the second highest Acc on the MoNuSeg dataset. The extracted context features from its encoder are fused with the multiscale features of the decoder through skip connections, which compensates for the loss of spatial information caused by downsampling. Other comparison methods are relatively weak in segmenting more precise areas.

It is apparent that UCTransNet presents great scores in both the QaTa-COV19 and Glas datasets. It designs Channel Transformer(CTrans) to replace the original skip connections, which helps solve the semantic gap and achieves accurate automatic medical image segmentation.
What is particularly striking about the quantitative results is that LViT introduces multimodal inputs but produces some suboptimal results. It may not make full use of its encoder to catch useful representations due to its end-to-end model.

With regard to the natural images, GTUNet and LViT gain the most on CoSKLE and MFFD, respectively. GTUNet is beneficial due to its grouping and bottleneck structure. LViT uses the vision-language strategy to achieve segmentation. 

\begin{figure*}[h!]
	\centering
	\begin{minipage}{0.05\linewidth}
		\centering
		\centerline{{\scriptsize \makecell{The left\\ infected region\\ is wider.}}}
	\end{minipage}
	\begin{minipage}{0.11\linewidth}
		\centering
		\includegraphics[width=0.55in,height=0.55in]{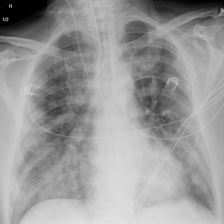}
		\label{IMG11}
	\end{minipage}
	\begin{minipage}{0.11\linewidth}
		\centering
		\includegraphics[width=0.55in,height=0.55in]{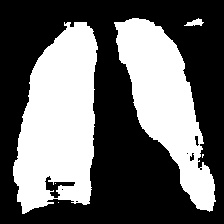}
		\label{NoDownViT1}
	\end{minipage}
	\begin{minipage}{0.11\linewidth}
		\centering
		\includegraphics[width=0.55in,height=0.55in]{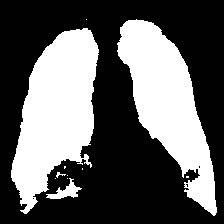}
		\label{NoUpViT1}
	\end{minipage}
	\begin{minipage}{0.11\linewidth}
		\centering
		\includegraphics[width=0.55in,height=0.55in]{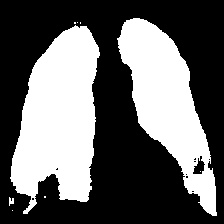}
		\label{NoSimSiam1}
	\end{minipage}
	\begin{minipage}{0.11\linewidth}
		\centering
		\includegraphics[width=0.55in,height=0.55in]{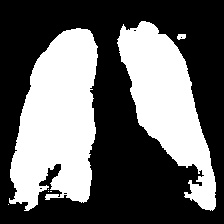}
		\label{NoPatchFuse1}
	\end{minipage}
	\begin{minipage}{0.11\linewidth}
		\centering
		\includegraphics[width=0.55in,height=0.55in]{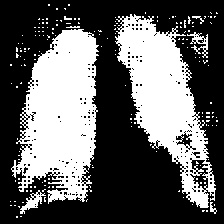}
		\label{NoUpAttention1}
	\end{minipage}
	\begin{minipage}{0.11\linewidth}
		\centering
		\includegraphics[width=0.55in,height=0.55in]{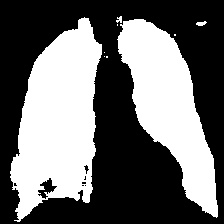}
		\label{OURS11}
	\end{minipage}
	\begin{minipage}{0.11\linewidth}
		\centering
		\includegraphics[width=0.55in,height=0.55in]{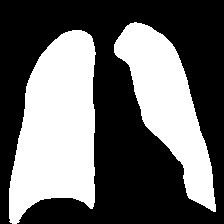}
		\label{GT11}
	\end{minipage}
	
	\vspace{0.1cm}
	\begin{minipage}{0.05\linewidth}
		\centering
		\centerline{{\scriptsize \makecell{The two\\ infected regions\\ are symmetric.}}}
	\end{minipage}
	\begin{minipage}{0.11\linewidth}
		\centering
		\includegraphics[width=0.55in,height=0.55in]{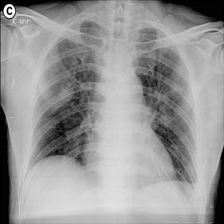}
		\label{IMG12}
	\end{minipage}
	\begin{minipage}{0.11\linewidth}
		\centering
		\includegraphics[width=0.55in,height=0.55in]{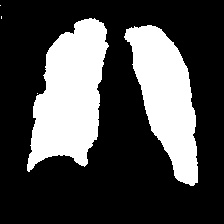}
		\label{NoDownViT2}
	\end{minipage}
	\begin{minipage}{0.11\linewidth}
		\centering
		\includegraphics[width=0.55in,height=0.55in]{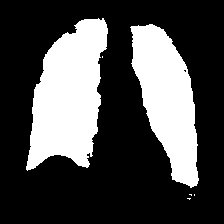}
		\label{NoUpViT2}
	\end{minipage}
	\begin{minipage}{0.11\linewidth}
		\centering
		\includegraphics[width=0.55in,height=0.55in]{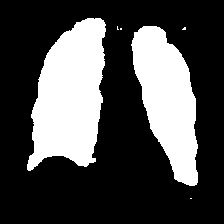}
		\label{NoSimSiam2}
	\end{minipage}
	\begin{minipage}{0.11\linewidth}
		\centering
		\includegraphics[width=0.55in,height=0.55in]{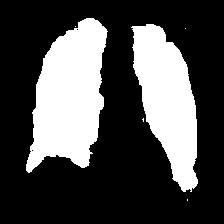}
		\label{NoPatchFuse2}
	\end{minipage}
	\begin{minipage}{0.11\linewidth}
		\centering
		\includegraphics[width=0.55in,height=0.55in]{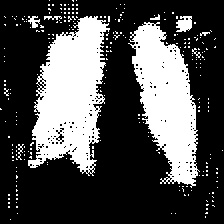}
		\label{NoUpAttention2}
	\end{minipage}
	\begin{minipage}{0.11\linewidth}
		\centering
		\includegraphics[width=0.55in,height=0.55in]{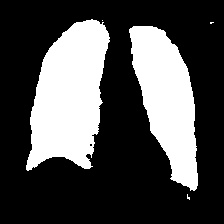}
		\label{OURS12}
	\end{minipage}
	\begin{minipage}{0.11\linewidth}
		\centering
		\includegraphics[width=0.55in,height=0.55in]{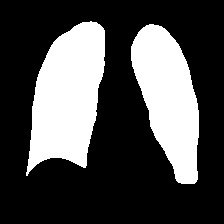}
		\label{GT12}
	\end{minipage}
	
	\vspace{0.1cm}
	\hspace{-0.1cm}
	\begin{minipage}{0.05\linewidth}
		\centering
		\vspace{-0.7cm}
		\centerline{{\scriptsize \makecell{The two\\ infected regions\\ are symmetric.}}}
	\end{minipage}
	\begin{minipage}{0.11\linewidth}
		\centering
		\includegraphics[width=0.55in,height=0.55in]{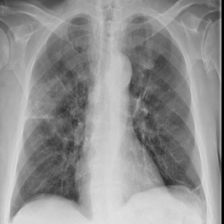}
		\centerline{\small  IMG}
		\label{IMG13}
	\end{minipage}
	\begin{minipage}{0.11\linewidth}
		\centering
		\includegraphics[width=0.55in,height=0.55in]{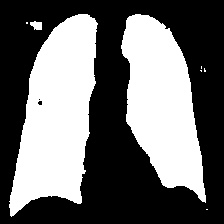}
		\centerline{\small  $\dagger$DownViT}
		\label{NoDownViT3}
	\end{minipage}
	\begin{minipage}{0.11\linewidth}
		\centering
		\includegraphics[width=0.55in,height=0.55in]{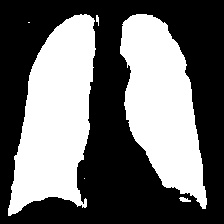}
		\centerline{\small  $\dagger$UpViT}
		\label{NoUpViT3}
	\end{minipage}
	\begin{minipage}{0.11\linewidth}
		\centering
		\includegraphics[width=0.55in,height=0.55in]{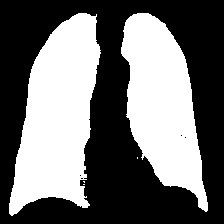}
		\centerline{\small  $\dagger$PPE}
		\label{NoSimSiam3}
	\end{minipage}
	\begin{minipage}{0.11\linewidth}
		\centering
		\includegraphics[width=0.55in,height=0.55in]{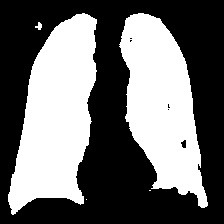}
		\centerline{\small  $\dagger$MSFF}
		\label{NoPatchFuse3}
	\end{minipage}
	\begin{minipage}{0.11\linewidth}
		\centering
		\includegraphics[width=0.55in,height=0.55in]{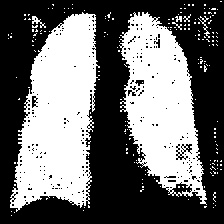}
		\centerline{\small  $\dagger$UpAttention}
		\label{NoUpAttention3}
	\end{minipage}
	\begin{minipage}{0.11\linewidth}
		\centering
		\includegraphics[width=0.55in,height=0.55in]{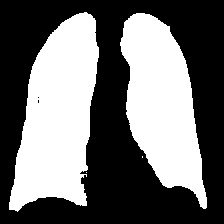}
		\centerline{\small  OURS}
		\label{OURS13}
	\end{minipage}
	\begin{minipage}{0.11\linewidth}
		\centering
		\includegraphics[width=0.55in,height=0.55in]{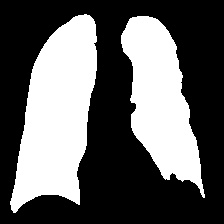}
		\centerline{\small  GT}
		\label{GT13}
	\end{minipage}
	\vspace{-0.4cm}
	\caption{Ablation Results of Medical image segmentation from QaTa-COV19 dataset.  $\dagger$denotes our model without this part.}
	\vspace{-0.4cm}
	\label{ablation}
\end{figure*}

\textbf{Precision and Recall:} Precision refers to the proportion of true positives identified as positive categories in a sample. Recall, on the other hand, measures the fraction of relevant instances that were retrieved. Both precision and recall are crucial metrics for evaluating the performance of a model. As shown in Tab. \ref{Quantitative result} and Tab. \ref{Natural result}, our model outperforms other methods in terms of precision and recall. These higher values indicate that our model can more accurately segment image regions, resulting in a more reliable output.

Our model achieves improvements to a certain extent in terms of these two metrics. This illustrates that the proposed model benefits from the 2-stage strategy. Moreover, MSFF and UpAttention are used to refine the representations from $PPE^{*}$. Therefore, we can conclude that it performs well with false positives and negatives.

Our model takes full advantage of the text prior prompts. PPE is adept in catching the representation of the input statistics. In addition, the hybrid Transformer and CNN reconcile the long-range dependencies and local features. This design can better merge text information and encode global features with Transformer while retaining the CNN's ability to excavate the local features from the images. Moreover, MSFF generates progressive features that further increases the mask accuracy.

\begin{table}[!h]
	\caption{Ablation experiments conducted on test dataset to study the effectiveness of proposed parts.}
	\centering
	\resizebox{\linewidth}{!}{
		\begin{tabular}[H]{ccccc|c|c|c}
			\hline  \bm{$DownViT$} &\bm{$UpViT$}&\bm{$PPE^{*}$} &\bm{$MSFF$} &\bm{$UpAttention$}  &\bm{$Dice$}&\bm{$mIoU$} &\bm{$Acc$}\\ \hline
			\XSolidBrush &\Checkmark  &\Checkmark&\Checkmark&\Checkmark  &91.29 &84.20 &93.93\\
			\Checkmark & \XSolidBrush &\Checkmark&\Checkmark&\Checkmark  &91.22 &84.01 &93.96\\
			\Checkmark & \Checkmark &\XSolidBrush&\Checkmark&\Checkmark  &90.91 &83.60 &93.66\\
			\Checkmark & \Checkmark &\Checkmark&\XSolidBrush&\Checkmark  &84.72 &73.78 &87.34\\
			\Checkmark & \Checkmark &\Checkmark&\Checkmark&\XSolidBrush  &91.19 &84.07 &93.77\\
			\Checkmark & \Checkmark &\Checkmark&\Checkmark&\Checkmark    &\textbf{91.53} &\textbf{85.67} & \textbf{95.01}\\
			\hline
			\multicolumn{8}{l}{\small \XSolidBrush of \bm{$PPE^{*}$} denotes our model in $Stage 2$, without the parameters from $Stage 1$.}
		\end{tabular}
	}
	\label{ablation2}
\end{table}

\subsection{Ablation Study}
We conduct a series of ablation experiments on the QaTa-COV19 dataset to verify the effectiveness of each part in our model. We use the three indices mentioned above for comparison, and the results are presented in Fig. \ref{ablation} and Tab. \ref{ablation2}.
First, we evaluate our model without DownViT or UpViT in U-shaped PPE. As shown in Fig. \ref{ablation}, without DownViT or UpViT, the model cannot capture the overall features. This causes partial deletion of the segmented organ. In addition, when we remove DownViT or UpViT separately, the three indices all decline compared with the complete model. When lacking the pretrained parameters from PPE on CL, we observe that the model is weak in depicting the medical image features, especially the edge shape. The decrease in the three indices is larger than that in the first two ablation experiments.

\begin{figure}[!h]
	\centering
	\includegraphics[width=0.7\linewidth]{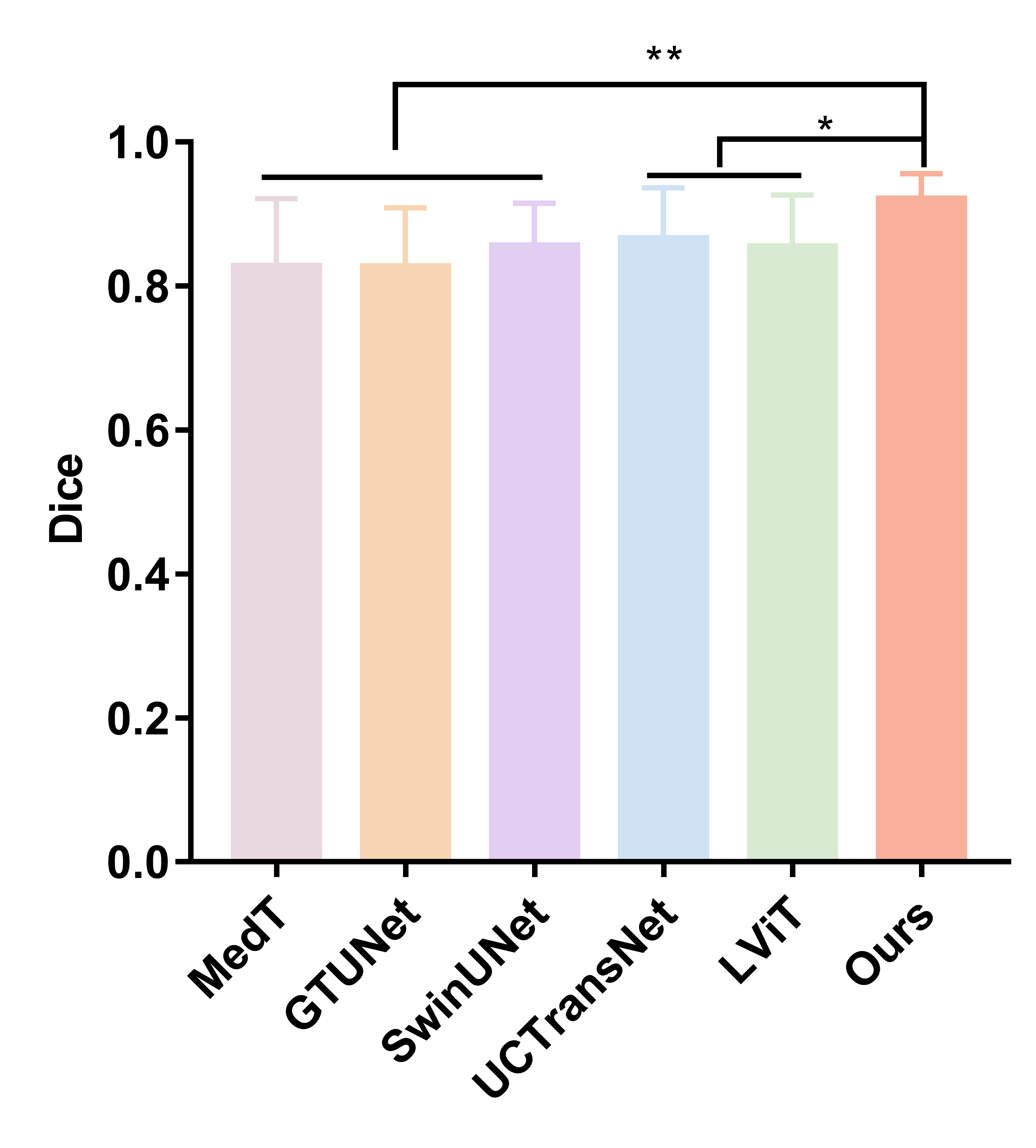}
	\caption{The t-test experiments about Dice on three medical datasets. All data are means ± standard deviation(SD). *p < 0.05; **p < 0.01; ***p < 0.001}
	\label{t-test-dice}
	\vspace{-0.3cm}
\end{figure}
\begin{figure}[!h]
	\centering
	\includegraphics[width=0.7\linewidth]{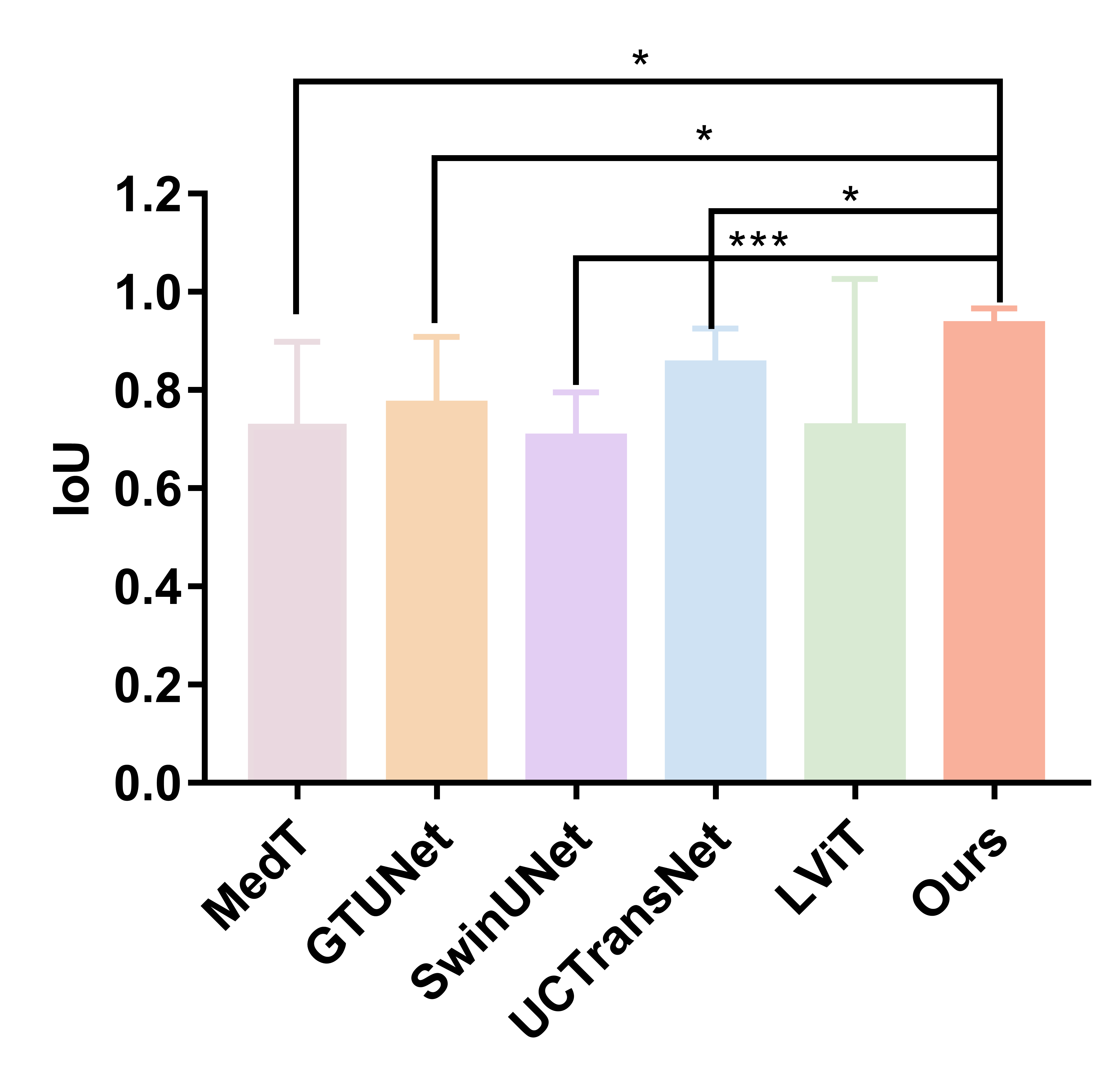}
	\caption{The t-test experiments about IoU on three medical datasets.}
	\label{t-test-iou}
	\vspace{-0.3cm}
\end{figure}

We also test the effect of the MSFF block and UpAttention block. Closer inspection of Fig. \ref{ablation} demonstrates that MSFF is important for integrating the holistic feature statistics for the final segmentation. Without this block, the model is deficient in handling the semantic gap between the natural data and medical data. In addition, the Dice value decreases from $91.53\%$ to $84,72\%$, which is the maximum. As illustrated in Fig. \ref{ablation}, the visual outcome of the model without UpAttention is subpar. As a result, the last step to refine the prediction mask is significant, and this part in our model is well designed.

\subsection{T-Test}
The t-test is a statistical hypothesis test that employs the Student's t-distribution when analyzing data under the null hypothesis. It is commonly used to determine if there is a significant difference between two sets of data. In our study, we conducted experiments using the t-test to compare Dice and IoU scores across three medical datasets (MoNuSeg, QaTa-COV19, and Glas). We employed GraphPad Prism Software version 6.0 for all statistical analyses. For comparisons among different segmentation methods, we utilized one-way ANOVA analysis. Our results showed that our model achieved significant improvements in both Dice and IoU scores. The significance threshold was set at $P < 0.05$. Overall, our findings support the effectiveness of our proposed approach in improving medical image segmentation performance.

As shown in Fig. \ref{t-test-dice}, our model demonstrates the most stable performance across various datasets. In comparison to UCTransNet and LViT, our model has achieved significant improvements. Additionally, the two stars obtained by comparing MedT, GTUNet, and SwinUNet indicate that our model has greatly advanced in Dice. Similarly, our model also achieved significant improvements when compared with other methods such as MedT, GTUNet, and UCTransNet in Fig. \ref{t-test-iou}. The model's segmentation ability was significantly improved, which is reflected by the three stars received when comparing it to SwinUNet. Although there was no significant improvement when comparing our model to LViT, its IoU value still outperformed that of the method. These results further demonstrate the superiority of our model.

\section{Conclusion}
In this paper, we propose text prior prompts to guide segmentation and de
vised a two-stage learning pipeline. In Stage 1, we pre-train a dual-input Transformer encoder called PPE on extensive natural datasets using contrastive learning. The U-shaped Transformer and CNN are integrated into PPE to balance local and nonlocal features. In Stage 2, we design a multi-scale feature fusion block (MSFF) and an UpAttention block. The MSFF is applied to develop multi-modality features obtained from PPE into multi-scale features, which addresses the semantic gap between natural data and medical data. The UpAttention block refines the predicted results by retaining the local features of the image. Our model's design encourages an efficient and accurate way to implement image segmentation, which outperforms other related methods on both medical and natural datasets.

%%Vancouver style references.
\bibliographystyle{cag-num-names}
\bibliography{cag-template}

\end{document}